\documentclass[11pt,a4paper]{article}
\pdfoutput=1 

\usepackage{jcappub} 

\usepackage[T1]{fontenc} 

\usepackage{graphicx}
\usepackage{epsfig}
\usepackage{amsmath}
\usepackage{amssymb}
\usepackage{subfigure}
\usepackage{longtable}
\usepackage{graphicx}
\usepackage{arydshln}
\usepackage{textcomp}
\usepackage{color}
\usepackage{url}
\usepackage{relsize}

\title{\boldmath Another look at distortions of the Cosmic Microwave Background spectrum}

   \author[a,b,1]{G. De Zotti, \note{Corresponding author.}}
   \author[a,c]{M. Negrello,}
   \author[b]{G. Castex,}
   \author[b,d]{A. Lapi}
   \author[e]{and M. Bonato}

\affiliation[a]{INAF-Osservatorio Astronomico di Padova, Vicolo dell'Osservatorio 5, I-35122 Padova, Italy}
\affiliation[b]{SISSA, Via Bonomea 265, 34136, Trieste, Italy}
\affiliation[c]{School of Physics \& Astronomy, Cardiff University, Queens Buildings, The Parade, Cardiff CF24 3AA, UK}
\affiliation[d]{Dipartimento di Fisica, Universit\`a ``Tor Vergata'', Via della Ricerca Scientifica 1, I-00133 Roma, Italy}
\affiliation[e]{Department of Physics and Astronomy, Tufts University, 574 Boston Avenue, Medford, MA 02155, USA}
\emailAdd{gianfranco.dezotti@oapd.inaf.it}
\emailAdd{NegrelloM@cardiff.ac.uk}
\emailAdd{gcastex@sissa.it}
\emailAdd{lapi@sissa.it}
\emailAdd{matteo.bonato@tufts.edu}
\emailAdd{danese@sissa.it}

\abstract{We review aspects of Cosmic Microwave Background (CMB) spectral distortions which do not appear to have been fully explored in the literature. In particular, implications of recent evidences of heating of the intergalactic medium (IGM) by feedback from active galactic nuclei are investigated. Taking also into account the IGM heating associated to structure formation, we argue that values of the $y$ parameter of $\hbox{several}\times 10^{-6}$, i.e. a factor of a few below the COBE/FIRAS upper limit, are to be expected. The Compton scattering by the re-ionized plasma also re-processes primordial distortions, adding a $y$--type contribution. Hence no pure Bose-Einstein-like distortions are to be expected. An assessment of Galactic and extragalactic foregrounds, taking into account the latest results from the \textit{Planck} satellite as well as the contributions from the strong C{\sc ii} and CO lines from star-forming galaxies, demonstrates that a foreground subtraction accurate enough to fully exploit the PIXIE sensitivity will be extremely challenging. Motivated by this fact we also discuss methods to detect spectral distortions not requiring absolute measurements and show that accurate determinations of the frequency spectrum of the CMB dipole amplitude may substantially improve over COBE/FIRAS limits on distortion parameters. Such improvements may be at reach of next generation CMB anisotropy experiments. The estimated amplitude of the Cosmic Infrared Background (CIB) dipole might be detectable by careful analyses of \textit{Planck} maps at the highest frequencies. Thus \textit{Planck} might provide interesting constraints on the CIB intensity, currently known with a $\simeq 30\%$ uncertainty.  }


\notoc

\begin{document}

\newcommand{\todo}[1]{\textsf{[TODO: #1]}}
\maketitle
\flushbottom

\def\simlt{\mathrel{\rlap{\lower 3pt\hbox{$\sim$}}\raise 2.0pt\hbox{$<$}}}
\def\simgt{\mathrel{\rlap{\lower 3pt\hbox{$\sim$}} \raise
2.0pt\hbox{$>$}}}
\def\lsim{\,\lower2truept\hbox{${<\atop\hbox{\raise4truept\hbox{$\sim$}}}$}\,}
\def\gsim{\,\lower2truept\hbox{${> \atop\hbox{\raise4truept\hbox{$\sim$}}}$}\,}
\def\aap{A\&A}
\def\apj{ApJ}
\def\apjs{ApJS}
\def\apjl{ApJL}
\def\mnras{MNRAS}
\def\aj{AJ}
\def\nat{Nature}
\def\aaps{A\&A Supp.}
\def\aaps{A\&A Supp.}
\def\pra{Phys.Rev.A}         
\def\physrep{Physics Reports}         
\def\prb{Phys.Rev.B}         
\def\prc{Phys.Rev.C}         
\def\prd{Phys.Rev.D}         
\def\prl{Phys.Rev.Lett}      
\def\araa{ARA\&A}       
\def\gca{GeCoA}         
\def\pasp{PASP}              
\def\pasj{PASJ}              
\def\apss{ApSS}
\def\jcap{JCAP}
\def\sovast{Soviet Astronomy}
\def\na{New Astronomy}
\def\aapr{A\&A Rev.}
\def\planss{Planet. Space Sci.}
 \vspace{1em}
\section{Introduction}\label{sect:intro}

Fifty years after the report of the discovery of the Cosmic Microwave Background \citep[CMB;][]{PenziasWilson1965,Dicke1965} and 25 years after the first report of CMB spectral measurements by the Far Infrared Absolute Spectrophotometer (FIRAS) on the Cosmic Background Explorer (COBE) satellite \citep{Mather1990}, deviations from a perfect black-body have still to be detected. Yet, since the Universe is obviously not in thermal equilibrium, the CMB spectrum cannot be exactly Planckian. Unavoidable CMB spectral distortions have been discussed by \citep{ChlubaSunyaev2012,SunyaevKhatri2013,Tashiro2014}. Some of them, which carry information on the cosmic history inaccessible by any other means, must be present at a level attainable by next generation instruments. In particular, the proposed Primordial Inflation Explorer \citep[PIXIE;][]{Kogut2014} promises improvements in measurement accuracy by 2--3 orders of magnitude compared to COBE/FIRAS.

The CMB spectral distortions that may be expected as the result of processes occurring in the early universe have been characterized primarily by Zeldovich and Sunyaev \citep{ZeldovichSunyaev1969,SunyaevZeldovich1970,Zeldovich1972,Sunyaev1974,IllarionovSunyaev1975a, IllarionovSunyaev1975b}. The evolution of distorted spectra and constraints on processes that can produce them have been investigated, both analytically and numerically, by numerous subsequent studies \citep{ChanJones1975a,ChanJones1975b,ChanJones1975c,ChanJones1975d,Dubrovich1975, DaneseDeZotti1977, DaneseDeZotti1978,DaneseDeZotti1980a,DaneseDeZotti1980b,SunyaevZeldovich1980,DaneseDeZotti1982,LyubarskySunyaev1983,KawasakiSato1986, KawasakiSato1987,Freese1987, Danese1990,BernsteinDodelson1990,Dorosheva1990,Burigana1991a,Burigana1991b,BarrowColes1991,Daly1991,Burigana1993, Ellis1992,HuSilk1993a,HuSilk1993b,Hu1994,HuSugiyama1994,Burigana1995, Refregier2000, Jedamzik2000, McDonald2001, SalvaterraBurigana2002,Oh2003, Burigana2004, ZizzoBurigana2005, BuriganaZizzo2006, RubinoMartin2006, RubinoMartin2008, ChlubaSunyaev2008, ChlubaSunyaev2009,  ProcopioBurigana2009, Carr2010, Burigana2011, Lochan2012, Dent2012, Chluba2012, ChlubaSunyaev2012, ChlubaErickcekBenDayan2012, Khatri2012a, Khatri2012b, KhatriSunyaev2012a, KhatriSunyaev2012b, KhatriSunyaev2013, Tashiro2013, Chluba2013a, Chluba2013b, Chluba2014a, Chluba2014b, Chluba2014c, AminGrin2014, Tashiro2014, KunzeKomatsu2014, Chluba2015, Desjacques2015}.

The prospect of CMB spectral measurements with a precision orders of magnitude higher than COBE/FIRAS, foreseen by the roadmap for the future NASA Astrophysics missions\footnote{http://science.nasa.gov/media/medialibrary/2013/12/20/secure-Astrophysics\_Roadmap\_2013.pdf}, has rekindled interest in the complete characterization of spectral distortions that may be produced during the cosmic history. In this paper, after a short summary of the theory of CMB distortions (Sect.~\ref{sect:basics}) we present an up-to-date estimate of Galactic and extragalactic foregrounds that may constitute the main limitation to accurate spectral measurements (Sect.~\ref{sect:foregrounds}). Next we discuss the distortions from the cosmic re-ionization (Sect.~\ref{sect:reion}), focussing in particular on an aspect which does not appear to have been fully explored in the literature, namely  the effect of mechanical energy released by Active Galactic Nuclei (AGNs; Sect.~\ref{sect:reionization}). Such energy release,  advocated by current models of galaxy-AGN co-evolution, may have had a major impact on the heating of the intergalactic medium. In Sect.~\ref{sect:dipole} we examine alternative methods to detect spectral distortions, that do not need absolute calibration, i.e. measurements of the CMB dipole spectrum and measurements of the Sunyaev-Zeldovich effect towards rich clusters of galaxies. The main conclusions are summarized in Sect.~\ref{sect:conclusions}.

We adopt a flat $\Lambda$CDM cosmology with the values of the parameters in the last column of Table~4 of \citep{PlanckParameters2015}, slightly rounded, i.e. $\Omega_\Lambda=0.69$, $\Omega_m=0.31$, $h=0.677$, $\Omega_b=0.049$, and a present-day CMB temperature $T_{\rm CMB}=2.725\,$K \citep{FixsenMather2002}.

\begin{figure*}
\includegraphics[width=0.48\textwidth, angle=0]{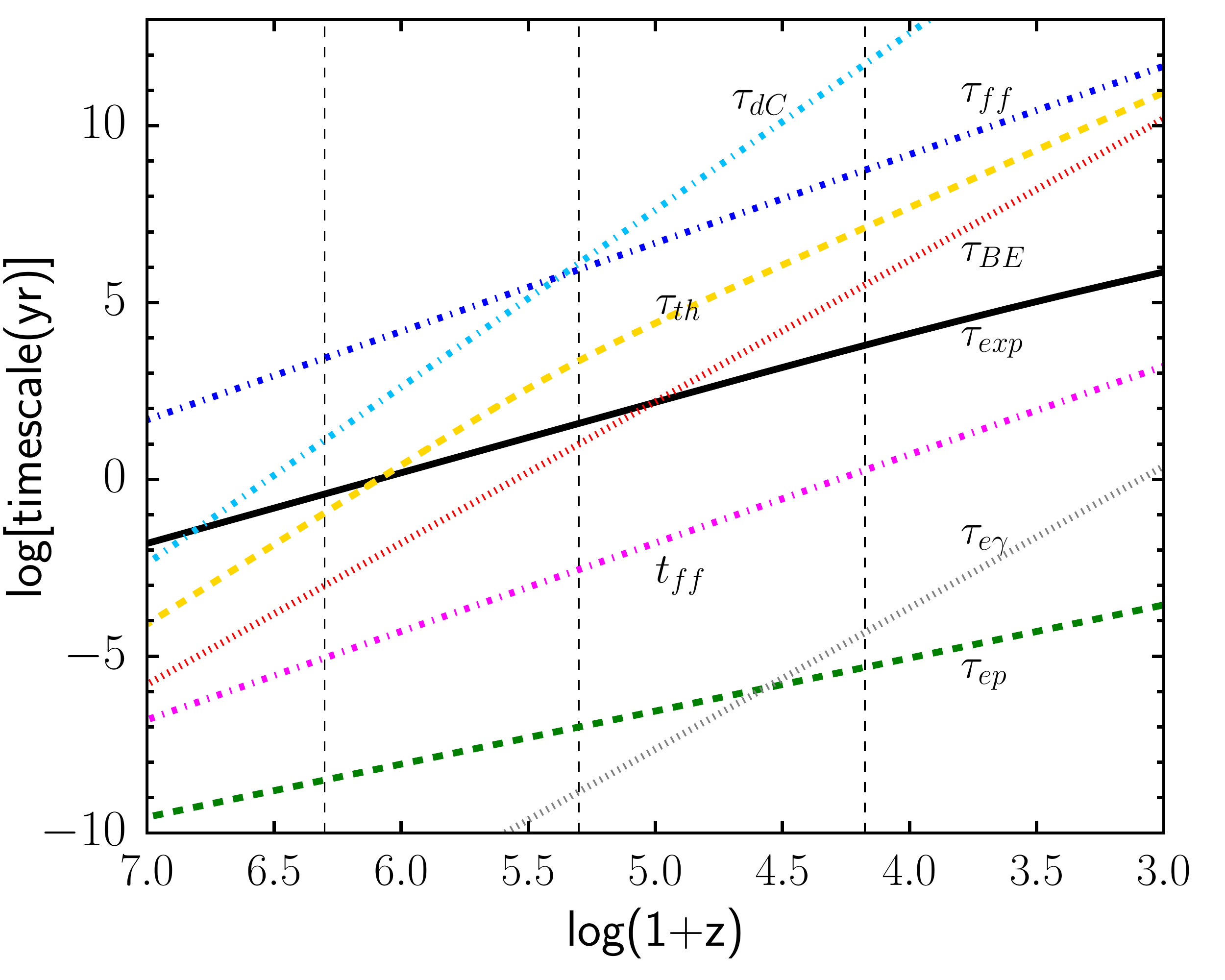}
\includegraphics[width=0.48\textwidth, angle=0]{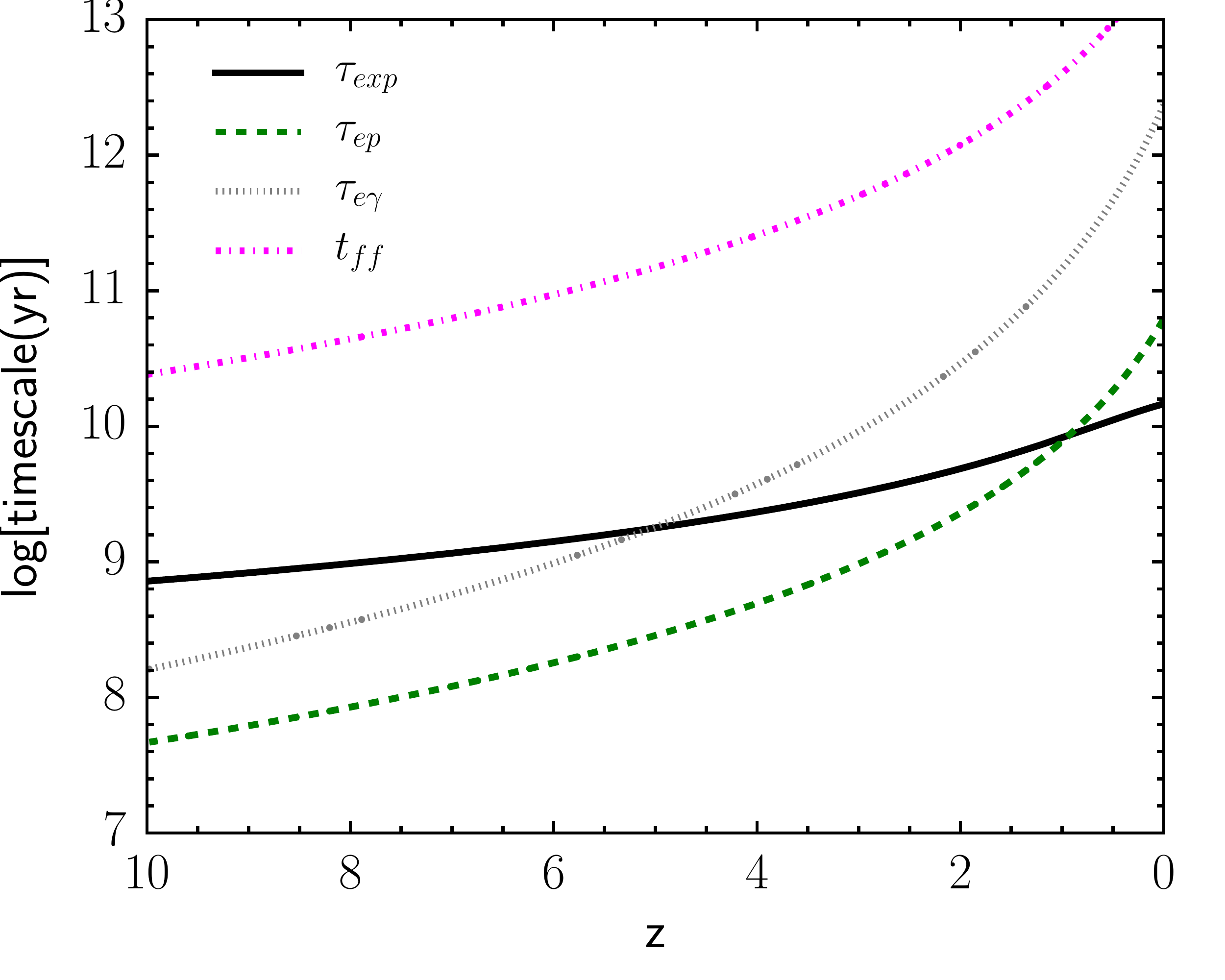}
\caption{\textbf{Left-hand panel:} timescales of processes governing the evolution of CMB distortions compared with the expansion timescale (thick black solid line) before recombination; $t_{\rm ff}$ was computed setting the clumping factor $C=1$. \textbf{Right-hand panel:} timescales for thermal equilibrium between electrons and the radiation field and between electrons and protons in the post-reionization era; $\tau_{\rm ep}$ was computed adopting the cosmic number density of baryons and an electron temperature $T_e=10^7\,$K; $t_{\rm ff}$ was computed setting $C=3$.}
 \label{fig:timescales}
\end{figure*}

\section{Basic theory}\label{sect:basics}

\subsection{Time scales}\label{sect:timescales}
Useful insights into the processes governing the evolution of the CMB spectrum come from consideration of interaction timescales, compared with
the cosmic expansion timescale, $\tau_{\rm exp}$ (see the left-hand panel of Fig.~\ref{fig:timescales}). For a flat $\Lambda$CDM universe we have:
\begin{equation}\label{eq:texp}
\tau_{\rm exp}={a(t)\over \dot{a}(t)}\simeq {4.56\times 10^{17}\over \left[\Omega_m (1+z)^3+\Omega_r(1+z)^4+\Omega_\Lambda\right]^{1/2}} \ \hbox{s},
\end{equation}
where $\Omega_r=1.69\, a_{\rm BB}\, T_{\rm CMB}^4/\rho_{\rm cr}\simeq 9.05\times 10^{-5}$, $a_{\rm BB}=7.56\times 10^{-15}\,\hbox{erg}\,\hbox{cm}^{-3}\,\hbox{K}^{-4}$ being the black-body constant and $\rho_{\rm cr}=8.61\times 10^{-30}\,\hbox{g}\,\hbox{cm}^{-3}$. The factor 1.69 takes into account the contribution of relativistic neutrinos according to the Standard Model (with an effective number of neutrino species $N_{\rm eff} = 3.046$, including a small correction caused by a non-thermal distortion of the spectra during electron-positron annihilation).

In the pre-recombination plasma, the relaxation time for thermal equilibrium between electrons and protons, \citep[$\tau_{\rm ep}$,][]{Spitzer1962},
\begin{equation}\label{eq:tep}
\tau_{\rm ep}\simeq 2.8\times 10^8 \left({T_e\over T_r}\right)^{3/2}(1+z)^{-3/2} \ \hbox{s},
\end{equation}
is much shorter than $\tau_{\rm exp}$ so that the two constituents of the plasma are kept at the same temperature. Here $T_r=T_{\rm CMB}(1+z)$ and $T_e$ is the electron temperature.

The coupling of the plasma with the radiation field is governed by Compton (Thomson) scattering. The Compton cooling time of hot electrons,  $\tau_{\rm e \gamma}$, is
\begin{equation}\label{eq:te_gamma}
\tau_{\rm e \gamma}= {3 m_e c \over 4 \sigma_T \epsilon_r}\simeq 7.4\times 10^{19}(1+z)^{-4} \ \hbox{s},
\end{equation}
where $\sigma_T$ is the Thomson cross section and $\epsilon_r=a T_r^4$; $\tau_{\rm e \gamma}$ is much shorter than the free-free cooling time
\begin{equation}\label{eq:free-free}
t_{\rm ff}= {3 n_e k T_e \over w_{\rm ff}}\simeq 9.6\times 10^{17}C^{-1}\left(\bar{g}(T_e)\over 1.2\right)^{-1}T_e^{1/2}(z)(1+z)^{-3}\ \hbox{s},
\end{equation}
where
\begin{equation}\label{eq:ff_cooling}
w_{\rm ff}= 9.5\times 10^{-41}C \left(\bar{g}(T_e)\over 1.2\right)T_e^{1/2}(z)(1+z)^{6}\ \hbox{erg}\,\hbox{s}^{-1}\,\hbox{cm}^{-3},
\end{equation}
is the free-free energy loss rate by a fully ionized plasma with cosmological composition, $C=\langle n_e^2 \rangle/\langle n_e \rangle^2$ is the clumping factor ($\langle ... \rangle$ denotes average values) and  $\bar{g}(T_e)$ is the frequency average of the velocity averaged Gaunt factor, whose values are in the range 1.1 to 1.5 \citep{RybickiLightman1979}. Before recombination baryon fluctuations are very small, so that $C\simeq 1$. In the re-ionization era, fluctuations in the inter-galactic medium are much larger but their amplitude is still debated. We adopt $C=3$ \citep{KuhlenFaucher2012}.

Before recombination $\tau_{\rm e \gamma}$ is also much shorter than $\tau_{\rm exp}$. The very large dimensionless specific photon entropy (photon entropy per electron divided by the Boltzmann constant $k$)
\begin{equation}\label{eq:s}
s= {4a_{\rm bb} T_{\rm CMB}^3\over 3n_e k }\simeq 6.72\times 10^9,
\end{equation}
ensures that the radiation acts as a powerful thermostat, keeping electrons and protons at the equilibrium temperature with the radiation field \citep{Peyraud1968,ZeldovichLevich1970}.

The situation is different in the reionization era ($z< z_{\rm reion}$). As illustrated by the right-hand panel of Fig.~\ref{fig:timescales}, $\tau_{\rm e \gamma}$ exceeds $\tau_{\rm exp}$ for $z\le 5$. Thus keeping the universe fully ionized at $z\gg 5$ is increasingly demanding in terms of energetics. For realistic values of the clumping factor \citep[$C\sim 3$,][]{KuhlenFaucher2012}, the cooling by free-free is always sub-dominant.

After reionization the plasma can be heated to very hight temperatures (see Sect.~\ref{sect:reionization}) and $\tau_{\rm ep}$
\begin{equation}\label{eq:tep_lowz}
\tau_{\rm ep}\simeq 1.95\times 10^{18} \left({T_e\over 10^7\,{\rm K}}\right)^{3/2}(1+z)^{-3} \ \hbox{s},
\end{equation}
can become comparable to, or even larger than $\tau_{\rm exp}$ for $z$ of a few units (right-hand panel of Fig.~\ref{fig:timescales}). Thus, electrons and protons could have different temperatures at very late times.

\subsection{Bose-Einstein ($\mu$-type) distortions}

In the Thomson limit, the photon-electron collision time is
\begin{equation}\label{eq:tgamma_e}
\tau_{\rm \gamma e}= (n_e \sigma_T c)^{-1}\simeq 2.3\times 10^{20}(1+z)^{-3} \ \hbox{s}.
\end{equation}
The mean energy gain of a photon by inverse Compton scattering off hot electrons is $kT_e/m_e c^2$ and the characteristic time for the establishment of a quasi-equilibrium spectrum by effect of the inverse Compton scattering is:
\begin{equation}\label{eq:tBE}
\tau_{\rm BE}= \tau_{\gamma\rm e}\left({kT_e\over m_e c^2}\right)^{-1}\simeq 5\times 10^{29}{T_r\over T_e}(1+z)^{-4} \ \hbox{s}.
\end{equation}
The quasi-stationary solution is a Bose-Einstein (BE) spectrum with photon occupation number \citep{SunyaevZeldovich1970}
\begin{equation}\label{eq:etaBE}
\eta_{\rm BE}= {1\over e^{x_e+\mu} -1},
\end{equation}
with
\begin{equation}\label{eq:xe}
x_e={h\nu\over kT_e} \simeq 1.761{\nu\over 100\rm GHz}{T_r\over T_e}
\end{equation}
so that $x_e=1$ corresponds to $\nu_e\simeq 56.786(T_e/T_r)$\ GHz.
The Compton equilibrium electron temperature in such radiation field is slightly higher than the radiation temperature: $T_e\simeq T_r(1+0.456\mu)$, for $\mu \ll 1$ \citep{IllarionovSunyaev1975a}. Hence, for distortions created by electron heating, the spectrum has a deficit of photons compared to a black-body spectrum at the equilibrium electron temperature\footnote{The situation is different for energy extraction (the adiabatic cooling effect) or for general photon injections. The effect of the latter processes  has been discussed in detail only recently \citep{Chluba2015c}.}.

\subsection{Thermalization of the CMB spectrum}

So far photon-producing processes have been neglected. However, although their electron cooling rate is small compared to inverse Compton, they do have an important role. The main photon-producing processes are the free-free and the double Compton scattering, whose characteristic timescales for the evolution of the photon occupation number are
\begin{equation}\label{eq:tff}
\tau_{\rm ff}= 4.8\times 10^{26}(1+z)^{-5/2}\ \hbox{s}
\end{equation}
and
\begin{equation}\label{eq:tdC}
\tau_{\rm dC}= 1.3\times 10^{40}(1+z)^{-5}\ \hbox{s}.
\end{equation}
The latter is therefore dominant at very high redshifts.

While the characteristic time for Compton diffusion is frequency-independent, the timescales for free-free or double Compton absorption are
\begin{equation}\label{eq:tabs}
\tau_{\rm abs\,ff,dC}=\tau_{\rm ff,dC}{x_e^3 \exp(x_e)\over g_{\rm ff,dC}[\exp(x_e)-1]},
\end{equation}
where $g_{\rm ff}$ and $g_{\rm dC}$ are the free-free and double Compton Gaunt factors that, assuming a pure hydrogen plasma, can be approximated by \citep{Draine2011}
\begin{equation}\label{eq:gaunt_ff}
g_{\rm ff}(\nu,{T_e})=\ln\left\{\exp\left[5.960-\frac{\sqrt{3}}{\pi}\ln\left({\nu\over \hbox{GHz}} \left({T_e\over 10^4 \hbox{K}}\right)^{-1.5}\right)\right]+\exp(1)\right\},
\end{equation}
and \citep{ChlubaSunyaev2012}
\begin{equation}\label{eq:gaunt_dC}
g_{\rm dC}(\nu,{T_e})=\left(1+ {3\over 2}x + {29\over 24}x^2 + {11\over 16}x^3 + {5\over 12}x^4\right)e^{-2x}.
\end{equation}
The photon absorption timescale decreases as $x_e^2$ for $x_e\ll 1$ and becomes shorter than $\tau_{\rm BE}$ for $x_e< x_c$ with
\begin{equation}\label{eq:xc}
x_{c}=\tau_{\rm BE}^{1/2}\left({g_{\rm ff}\over \tau_{\rm ff}} + {g_{\rm dC}\over \tau_{\rm dC}} \right)^{1/2}.
\end{equation}
For $x_e > x_{c}$ photons are scattered to higher frequencies before they can be absorbed while for $x_e < x_{c}$ photon absorption is faster and an equilibrium spectrum is established. The solution of the kinetic equation including both Compton scattering and photon emitting processes has a frequency dependent chemical potential
\begin{equation}\label{eq:mu_xe}
\mu(x_e)=\mu_0 e^{-x_c/x_e}.
\end{equation}
At early enough times the cooperation of photon-emitting processes, that mostly produce low-frequency photons, and of Compton scattering that moves them up in frequency, leads to complete thermalization of the distortions. The thermalization timescale is
\begin{equation}\label{eq:t_therm}
\tau_{\rm th}=\left(\tau_{\rm BE}\over 1/\tau_{\rm ff} + 1/\tau_{\rm dC} \right)^{1/2}.
\end{equation}
A rough approximation of the thermalization redshift is obtained setting $\tau_{\rm th}\simeq \tau_{\rm exp}$. The detailed study by \citep{KhatriSunyaev2012a} yielded $z_{\rm th}\simeq 2\times 10^6$ \citep[see also][]{DaneseDeZotti1982,Burigana1991a,HuSilk1993b}.

What about cyclotron emission? Possible imprints of a stochastic background of primordial magnetic fields (PMFs) on CMB anisotropies, on CMB polarization and on non-Gaussianities were investigated by \citep{Planck_mag_field2015} using \textit{Planck} data. The derived 95\% confidence upper limit on the present day comoving magnetic field value, $B_0$, at a scale of $1\,$Mpc assuming magnetic flux freezing [$B=B_0 (1+z)^2$] was found to be $B_{1\,\mathrm{Mpc}}< 5.6\,$nG \citep[see also][]{Chluba2015b}.

In the case of thermal electrons, the cyclotron energy loss rate is \citep{Tortia1961}:
\begin{equation}
w_{\rm cycl}={4\over 3}\, {e^4 B^2 n_e\over m_e^3 c^5}k T_e \simeq 8\times 10^{-47} \left({B_0\over 40\,\hbox{nG}}\right)^2 {T_e\over T_r}(1+z)^{8}\,\hbox{erg}\,\hbox{cm}^{-3}\,\hbox{s}^{-1}.
\end{equation}
Its ratio with the free-free energy loss rate [eq,~(\ref{eq:ff_cooling})] is
\begin{equation}
{w_{\rm ff}\over w_{\rm cycl}}\simeq 5.85\times 10^7 {C\over 3} \left(\bar{g}(T_e)\over 1.2\right)\left({T_e\over T_r}\right)^{-1/2} \left({B_0\over 5\,{\rm nG}}\right)^{-2} (1+z)^{-1.5},
\end{equation}
suggesting that, if $B$ is not too far from its upper limit, the cyclotron emission may be the dominant photon producing process at high $z$ \citep{Afshordi2002}. However \citep{ZizzoBurigana2005} showed that in fact cyclotron has a negligible impact on the evolution of spectral distortions. This is because photons are emitted at very low frequencies as primordial magnetic fields have an extremely low amplitude. The characteristic dimensionless frequency is
\begin{equation}
x_{\rm e, cycl}\simeq 2\times 10^{-12} \left({T_e\over T_r}\right)^{-1} \left({B_0\over 40\,{\rm nG}}\right) (1+z),
\end{equation}
much lower, for values of $B_0$ consistent with observational constraints, than the frequency $x_c$ at which an equilibrium spectrum is established by the free-free and double Compton processes ($x_c>5\times 10^{-3}$).

A rough estimate of the minimum redshift at which $\mu$-type distortions can be produced is obtained from $\tau_{\rm BE}\simeq \tau_{\rm exp}$, giving $z_{\rm BE,min}\simeq 10^{5}$ \citep{Hu1995,ChlubaSunyaev2009}. The exact value for $z_{\rm BE,min}$ depends on the criterion that is applied (e.g., at what precision one calls the distortion `pure $\mu$--type distortion'; \citep{Khatri2012a,Khatri2012b} give $z_{\rm BE,min} = 2\times 10^{5}$, while \citep{Chluba2013c} finds that it must be closer to $3\times 10^5$ to get a $\mu$--type spectrum at the 1\% precision.

%

\subsection{Weak comptonization ($y$-type) distortions}

When $\tau_{\rm exp}/\tau_{\rm BE}\simlt 0.01$  \citep[$z< z_y\simeq 10^4$;][]{Hu1995,ChlubaSunyaev2009,KhatriSunyaev2012a, Chluba2013c}, i.e. in the weak comptonization limit, the shape of the distorted spectrum is slightly different depending on whether there is direct electron heating or, for example, mixing of black-bodies of different temperatures. A simple discussion of CMB distortions due photon injection was presented by \citep{HuSilk1993b} while a thorough investigation was carried out by \citep{Chluba2015c}.

Again, there are differences in the estimates for $z_y$ reported in the literature, due to the differences in the criterion for considering the spectrum `pure $y$--type'. \citep{KhatriSunyaev2012a} give $z_y\simeq 1.5\times 10^4$, while \citep{Chluba2013c} finds $z_y\simeq 1\times 10^4$ to get a $y$--type spectrum at the 1\% precision. Setting:
\begin{equation}\label{eq:y}
y(t)= \int_{t_0}^t {{\rm d}t' \over t_{\rm BE}},
\end{equation}
and $x=h\nu/kT_r$, $T_r$ being the temperature of a black-body with the same photon number density as the actual radiation field, the Compton term of the Kompaneets \citep{Kompaneets1957} equation writes:
\begin{equation}\label{eq:Kom}
{\partial\eta \over \partial y}= {1\over x^2}{\partial \over \partial x}\left\{x^4\left[{\partial\eta \over \partial x}+{T_r\over T_e}\left(\eta(x)+\eta^2(x)\right)\right]\right\},
\end{equation}
where $T_e$ is the electron temperature. In the case of small distortions
\begin{equation}\label{eq:Planck}
\eta(x)\simeq \eta_{\rm P}(x)= {1\over e^x -1},
\end{equation}
so that
\begin{equation}\label{eq:dy}
{\partial\eta \over \partial x}\simeq -[\eta(x)+\eta^2(x)],
\end{equation}
and the Compton term can be approximated by the ``diffusion'' equation
\begin{equation}
\left({\partial \eta \over \partial u}\right)_C \simeq {1\over x^2} {\partial \over \partial x}\left[x^4 {\partial\eta \over \partial x}\right].
\end{equation}
with
\begin{equation}\label{eq:u}
u(t)= \int_{t_0}^t \left(1-{T_r\over T_e}\right) {{\rm d}t' \over t_{\rm BE}}.
\end{equation}
With a suitable change of variable \citep{DaneseDeZotti1977} this equation takes the form of the one-dimensional heat conduction whose solution is \citep{ZeldovichSunyaev1969}
\begin{equation}\label{eq:ydist}
\eta(x,u)= {1\over (4\pi u)^{1/2}} \int_0^\infty \eta_i(x',0)\exp\mathlarger\{-{1\over 4u}\ln^2\mathlarger[{x\over x'}\exp(3u)\mathlarger]\mathlarger\}{dx'\over x'} ,
\end{equation}
which holds for any initial radiation spectrum $\eta_i(x,0)$.
If $ux^2\ll 1$ and the initial spectrum is black-body, eq.~(\ref{eq:ydist}) simplifies to \citep{ZeldovichSunyaev1969}
\begin{equation}\label{eq:compt}
\eta_C(x,u)=\eta_{\rm P}(\tilde{x})\left[1+u{\tilde{x}\exp(\tilde{x})\over \exp(\tilde{x})-1}\left({\tilde{x}\over \tanh(\tilde{x}/2)} -2 \right)\right],
\end{equation}
where $\tilde{x}=x(T_{r}/T_{\rm RJ})$ with $T_{\rm RJ}=T_{\rm r}\exp(-2y)$ and $\tanh(x)=[\exp(2x)-1]/[\exp(2x)+1]$.

More generally, the distorted spectrum in the weak comptonization case can be described by a superposition of black-body spectra
\begin{equation}\label{eq:kernel}
\eta(x,y)=\int_0^\infty R(T,y)\eta_{\rm P}(\nu,T)\,{\rm d}T\, .
\end{equation}
The temperature distribution function depends on the specific heating process. One example, referring to electron heating, was shown above. But important processes that occur in the early universe, such as dissipation of adiabatic density perturbations, may result in distortions generated by intermixing of photons coming from plasma clouds having macroscopic relative velocities without direct electron heating. The temperature distribution seen by an external observer in the case of clouds with a Maxwellian velocity distribution, each containing radiation at the same temperature $T_c$, is \citep{Zeldovich1972}:
\begin{equation}\label{eq:R}
R(T,w)= {1\over T_c\sqrt{4\pi w}} \exp\left[-{1\over 4 w}\left({T\over T_c}-1\right)^2 \right],
\end{equation}
where $w=\langle\beta^2\rangle/6$, $\langle\beta^2\rangle$ being the velocity dispersion of the clouds in units of $c$.

The main difference among the spectra given by eq.~(\ref{eq:ydist}) and by eq.~(\ref{eq:kernel}) plus eq.~(\ref{eq:R}) is due to a temperature shift $T_c=T_r(1-2w)$, where $T_r$ is the temperature in the definition of $x$ in eq.~(\ref{eq:ydist})\footnote{We are indebted to Jens Chluba for pointing this out to us.}. Allowing for this shift, the 2 spectra are indistinguishable to first order in $y$ or $w$.

\subsection{Intermediate epoch distortions}

An early analysis of CMB spectral distortions generated after the BE epoch and before that of weak comptonization, i.e. in the redshift range $1.5\times 10^{4}\simlt z \simlt 2\times 10^5$, was carried out by \citep{Burigana1991a}. The first explicit calculation of the distortion produced by partial comptonization was performed by \citep{ChlubaSunyaev2012}. A more detailed discussion of the numerical solutions in the intermediate era was given by \citep{KhatriSunyaev2012b,KhatriSunyaev2013}. Additional clarifications and improved approximate descriptions of the distortion in this transition regime were given by \citep{Chluba2013c}.

These authors have shown that these intermediate epoch distortions provide much more information than $\mu$-- and $y$--type distortions. While the latter are informative mainly on the amount of energy injected into the CMB, the shape of the former depends sensitively on the redshift of energy injection and allows us to distinguish among the mechanisms that have been operating. For example, as first shown by \citep{ChlubaSunyaev2012}, it may be possible to distinguish between particle annihilation and Silk damping. Efficient methods for distinguishing sources of distortions were expounded by \citep{Chluba2013b,ChlubaJeong2014}.

\subsection{Sources of spectral distortions}

Spectral distortions produced by physical processes that must have occurred during the history of the Universe have been thoroughly reviewed by \citep{ChlubaSunyaev2012,SunyaevKhatri2013,Tashiro2014,Chluba2014b}. 
Here we only remind a point that, although not new, is less well known. Even in the absence of energy injections, small spectral distortions occur. When baryons are non-relativistic, their adiabatic index is $5/3$. This causes the baryon temperature to be proportional to $(1 + z)^2$ so that they  cool faster than CMB photons. But, as long as electrons are tightly coupled with the radiation field, photons keep heating up the baryons. Thus the radiation temperature evolves not as $T_r \propto (1+z)$ but approximately as $T_R\propto (1+z)^{1+\epsilon}$ \citep{DaneseDeZotti1977} with
\begin{equation}\label{eq:s}
\epsilon= {3n_e K \over 4a_{\rm bb} T_{\rm CMB}^3} = s^{-1}\simeq 1.49\times 10^{-10},
\end{equation}
$s$ being the dimensionless specific photon entropy [eq.~(\ref{eq:s})]. If $T_i$ is the CMB temperature at the minimum redshift, $z_i$, at which thermalization of spectral distortions is ensured and $T_r=T_i(1+z)/(1+z_i)$ is the blackbody temperature corresponding to the photon number density we have
\begin{equation}\label{eq:Tdev}
\Delta T \equiv T_r - T_R = T_i\left[{1+z\over 1+z_i}-\left({1+z\over 1+z_i}\right)^{1+\epsilon}\right] \simeq T_i\left[{1+z\over 1+z_i}- {1+z\over 1+z_i} -{1+z\over 1+z_i}\ln\left({1+z\over 1+z_i}\right)\epsilon\right],
\end{equation}
i.e. \citep{Chluba2005,ChlubaSunyaev2012}
\begin{equation}\label{eq:dTdev}
{\Delta T \over T_r}= \epsilon\,\ln\left({1+z\over 1+z_i}\right)
\end{equation}
and $\Delta E/E_r\simeq 4 (\Delta T/T_r)$. The photon cooling results in an \textit{excess} of photons compared to a perfectly black-body spectrum. Detailed calculations \citep{ChlubaSunyaev2012} have shown that the resulting spectrum has a $\mu$-type distortions at GHz frequencies and a $y$-type distortion at high frequencies but with \textit{negative} values of the parameters. The distortions are tiny ($\mu_0\sim -2\times 10^{-9}$, $y\sim -4.3\times 10^{-10}$).

Those of $y$-type are easily swamped by the much larger \textit{positive} distortions expected at low $z$. Those of $\mu$-type may partially or even entirely cancel the positive $\mu$ distortions created by dissipation of acoustic waves on small scales due to photon diffusion \citep{Khatri2012a}. However, as shown by \citep{Chluba2012} using the full description of the dissipation physics, the cancelation  effect strongly depends on the values of the spectral index of scalar perturbations, $n_S$, and of its possible running, $n_{\rm run}=\hbox{d}n_S/ \hbox{d}\ln k$. For the standard values, $n_S\simeq 0.96$ and $n_{\rm run}\simeq 0$, the cancelation is small.

The photon cooling via Compton scattering moves the excess photons to lower and lower frequencies. However, as shown by \citep{Khatri2012a}, this does not lead to a Bose--Einstein condensation because the process is slow enough that the condensed photons are efficiently absorbed by double Compton scattering and free-free.

\section{Foregrounds}\label{sect:foregrounds}

\begin{figure*}
\includegraphics[width=0.48\textwidth, angle=0]{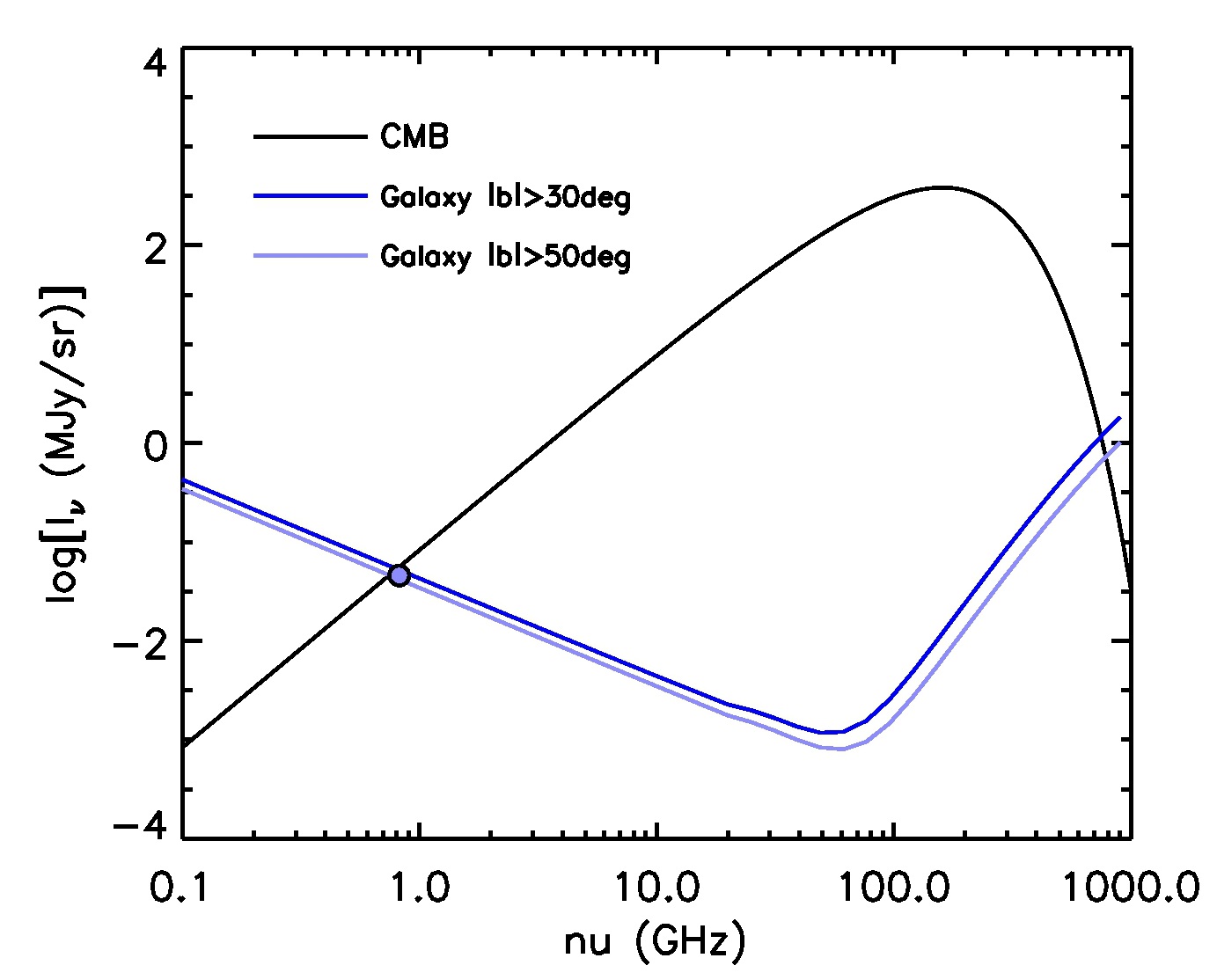}
\includegraphics[width=0.48\textwidth, angle=0]{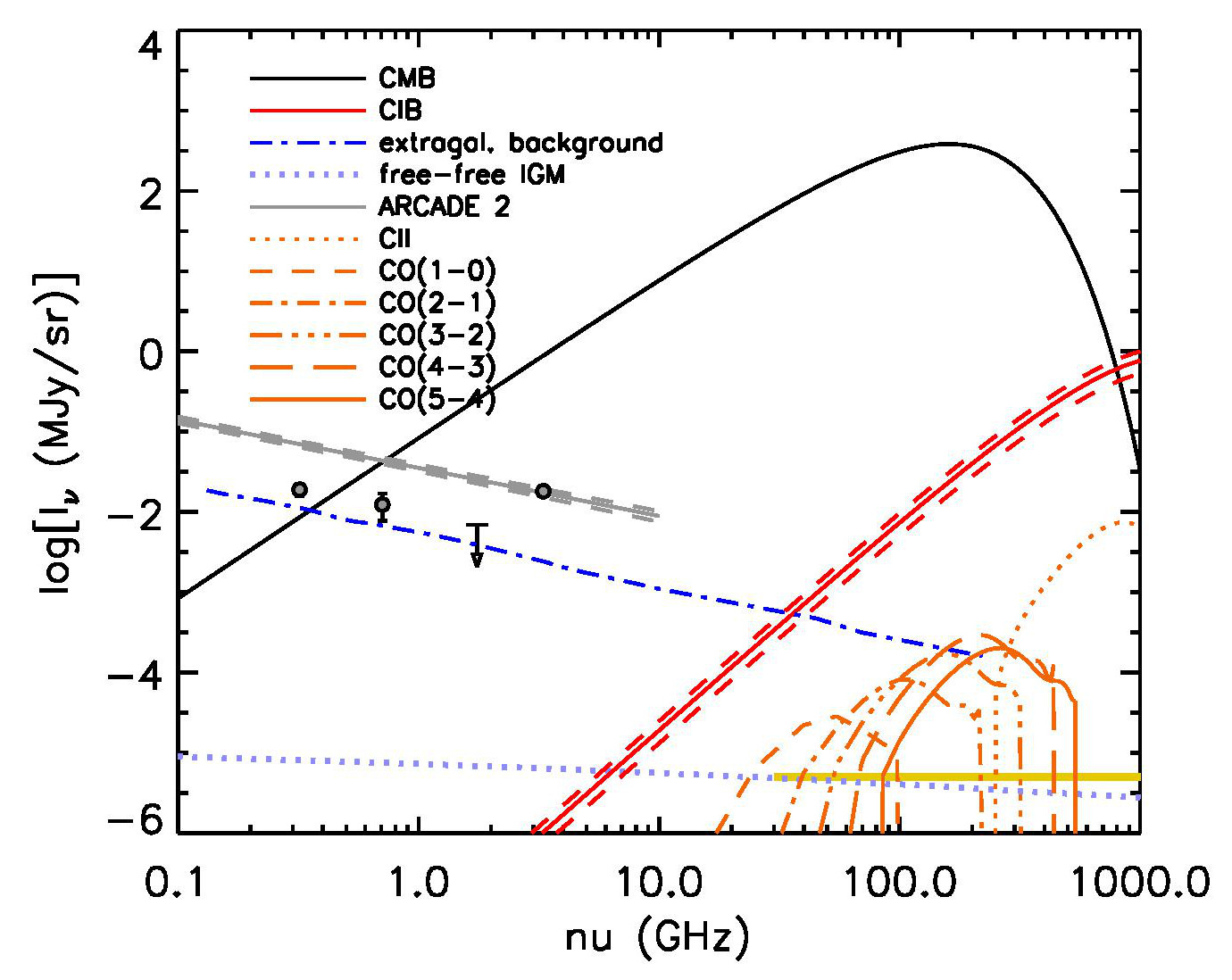}
\caption{Galactic and extragalactic foreground intensity compared with the CMB spectrum, assumed to be a perfect blackbody at $T=2.725\,$K. The \textbf{left-hand panel} shows the intensity of Galactic emissions averaged over latitudes $|b|>30^\circ$ (thick blue line) and $|b|>50^\circ$, based on an updated version of the Planck Sky Model \citep{Delabrouille2013}. The circle shows the measurement at 0.82\,GHz by \citep{Gervasi2008} in a region around $l=178.45^\circ$, $b=52.8^\circ$ ($T_{\rm gal}= 2.21\,$K; see the Appendix, Sect.~\ref{sect:conversion}, for the conversion from temperature to intensity). The \textbf{right-hand panel} shows the spectra of extragalactic foregrounds. The data points are observational estimates, all in terms of antenna temperature, by \citep{Wall1970}, $T_a=6\pm 1\,$K at 0.32 GHz and $T_a=0.8\pm 0.3\,$K at 0.707 GHz, and by \citep{Vernstrom2015} at 1.75\,GHz, $T_a=73 \pm 10\,$mK, claimed to be an upper limit \citep[see][for a review]{DeZotti2010}. The dot-dashed blue line is the radio background spectrum estimated using the models by  \citep{Massardi2010}, \citep{DeZotti2005} and \citep{Tucci2011} that accurately fit the radio source counts in different frequency ranges. The grey line above it represents the ARCADE~2 estimates \citep{Fixsen2011}, with uncertainties depicted by the dashed lines, based on their measurement at 3.3\,GHz (circle) and on a re-analysis of earlier data. The solid red line shows the spectrum of the Cosmic Infrared Background (CIB) determined by \citep{Fixsen1998} with the $1\,\sigma$ uncertainty represented by the dashed red lines. Also shown are estimates of contributions to the background of CO and C{\sc ii} lines from star-forming galaxies and of the free-free background produced at the re-ionization epoch (dotted line at the bottom of the figure; see sub-Sect.~\ref{sect:reionization}). The solid yellow line on the bottom right represents the PIXIE $1\,\sigma$ sensitivity \citep{Kogut2014}.
 }
 \label{fig:foregrounds}
\end{figure*}

Figure~\ref{fig:foregrounds} shows up-to-date estimates of Galactic and extragalactic foreground spectra. The average Galactic emissions at Galactic latitudes $|b|>30^\circ$ and $|b|>50^\circ$ have been obtained using an updated version of the Planck Sky Model \citep[PSM;][]{Delabrouille2013} that incorporates the recently published measurements by the \textit{Planck} satellite \citep{Planck_dust2014,Planck_comp_sep2014, Planck_therm_em2015, Planck_low_freq_diff2015}.

The PSM aims at providing a simulation of the sky as realistic as possible over a broad frequency range. It includes, in addition to a model of the CMB, Galactic diffuse emissions (synchrotron, free-free, thermal and spinning dust, CO lines), Galactic HII regions, extragalactic radio sources, dusty galaxies, the Cosmic Infrared Background (CIB), thermal and kinetic Sunyaev-Zeldovich signals from clusters of galaxies. Each component is simulated by means of educated interpolations/extrapolations of the available data, complemented by state-of-the-art models.

Distinctive features of the simulations are spatially varying spectral properties of synchrotron and dust; different spectral parameters for each point source; modelling of the clustering properties of extragalactic sources and of the power spectrum of CIB fluctuations, in close agreement with the latest observational determinations.

For the CIB spectrum we have used the analytical description by \citep{Fixsen1998}:
\begin{equation}\label{eq:CIB}
I_{\rm CIB}(\nu) = 1.734\times 10^{-4} T^3_{\rm CIB}\left({\nu \over \nu_0}\right)^{k_F} {x^3_{\rm CIB}\over \exp(x_{\rm CIB})-1}\,\hbox{MJy/sr},
\end{equation}
or
\begin{equation}\label{eq:nuCIB}
\nu I_{\rm CIB}(\nu) = 5.1998\times 10^{-3}\, T^3_{\rm CIB}\left({\nu \over \nu_0}\right)^{k_F+1} {x^3_{\rm CIB}\over \exp(x_{\rm CIB})-1}\,\hbox{nW}\,\hbox{m}^{-2}\,\hbox{sr}^{-1},
\end{equation}
with $T_{\rm CIB}=18.5\pm1.2\,$K, $\nu_0=3\times 10^{12}\,$Hz$, x_{\rm CIB}=h\nu/kT_{\rm CIB}=7.78(\nu/\nu_0)$  and $k_F= 0.64 \pm 0.12$. The uncertainties cannot be inferred from the errors on the parameters because the covariance matrix is not given. They were taken from the lower panel of Fig.~4 of \citep{Fixsen1998}.

As illustrated by Fig.~\ref{fig:foregrounds}, foreground emissions exceed the CMB below $\sim 0.8\,$GHz and above $\sim 800\,$GHz. Their intensity is always orders of magnitude above the PIXIE sensitivity, implying that a full exploitation of the potential of this experiment to pin down spectral distortions orders of magnitude weaker than the current COBE/FIRAS upper limits will require an extremely refined foreground subtraction.

In the PIXIE frequency range, the dominant extragalactic foreground is the CIB. Its subtraction will be complicated by spectral bumps due to the integrated emission of strong far/IR--mm lines such as the CO and the C{\sc ii}\,$157.7\,\mu$m lines produced by star-forming galaxies. Tentative estimates of the contributions of the strongest far-IR/sub-mm lines are shown in the right-hand panel of Fig.~\ref{fig:foregrounds}.

Such estimates were worked out exploiting a model for the cosmological evolution of the infrared (IR; 8--$1000\,\mu$m) luminosity function \citep{Cai2013}, fitting all the available observational determinations that extend up to $z\simeq 4$, coupled with relations between line intensities and IR luminosities, $L_{\rm IR}$. Data on the correlation between the C{\sc ii}\,$157.7\,\mu$m line intensity and $L_{\rm IR}$ have been discussed by \citep{Bonato2014} who found that they are consistent with a linear relation ($\log(L_{\rm C{\protect{\sc II}}}/L_{\rm IR})=-2.74$ with a dispersion of 0.37) except for local UltraLuminous IR Galaxies (ULIRGs) whose C{\sc ii} luminosities appear to be independent of $L_{\rm IR}$.  For the latter objects \citep{Bonato2014} found a mean $\log(L_{\rm C{\sc II}}/L_\odot)=8.85$ with a dispersion of 0.29.

As for the CO lines, we have used the $L_{\rm IR}$--CO luminosity relations ($\log L_{\rm IR} = \alpha \log L'_{\rm CO} +\beta$) for the CO rotational ladder from $J = 1-0$ to $J = 5-4$ with the best-fit slopes ($\alpha$) and intersection points ($\beta$) given in Table~3 of \citep{Greve2014}. The higher order CO transitions, for which relations with $L_{\rm IR}$ are also given by \citep{Greve2014}, have been neglected since their contributions are small compared to that of the C{\sc ii}\,$157.7\,\mu$m line. The $L'_{\rm CO}$ luminosities, in units of  $\hbox{K}\,\hbox{km}\,\hbox{s}^{-1}\hbox{pc}^2$, have been converted to luminosities in solar units, $L_{\rm CO}$, using the formula \citep{SolomonVandenBout2005}:
\begin{equation}
{L_{\rm CO}\over L_\odot}=3.18\times 10^{-11} \left({\nu\over {\rm GHz}}\right)^3 {L'_{\rm CO}\over \hbox{K}\,\hbox{km}\,\hbox{s}^{-1}\hbox{pc}^2}
\end{equation}
The right-hand panel of Fig.~\ref{fig:foregrounds} shows that the contributions of CO and C{\sc ii} lines to the CIB is $\sim 1\%$ from $\sim 40\,$GHz to $\sim 1\,$THz and substantially higher than the PIXIE sensitivity, especially at sub-mm wavelengths. Thus, the CIB is not completely spectrally smooth, a fact that complicates its subtraction. On the other hand, the PIXIE spectral coverage with many frequency channels is optimally suited for the foreground subtraction purpose. And the accurate measurements of the CIB spectrum that PIXIE can provide are important \textit{per se}. In fact, the current uncertainty on the CIB spectrum is one of the main limitations to a full understanding of the energetics of dust-obscured star formation and AGN accretion.

An independent estimate of the cumulative CO emission from galaxies has been worked out by \citep{Mashian2016}. Spectral distortions induced by a non-uniform distribution of C{\sc ii} emitting regions have been estimated by \citep{Pallottini2015}.

\section{Comptonization distortions from re-ionization}\label{sect:reion}

\subsection{Short historical notes}

Already very weak constraints on distortion parameters ($y\le 0.15$), combined with an upper limit of $1\,$K to the brightness temperature of the free-free emission from the inter-galactic plasma at 600\,MHz, allowed \citep{Sunyaev1968} and \citep{ZeldovichSunyaev1969} to infer ``The necessity of a neutral hydrogen period in the evolution of the universe'', provided that the cosmic baryon density is not too low. An absolute upper limit to the re-ionization redshifts was found to be $z_{\rm max}\simeq 300\Omega^{-7/9}$ or, distinguishing between $\Omega$ and $\Omega_b$ (at the time the dark matter, and even less the dark energy, were not yet within the astrophysicists' horizon),
\begin{equation}
z_{\rm max}< 300\left({\delta I_{\rm ff}(600\,{\rm MHz})\over 10^{-2}{\rm MJy}\,{\rm sr}^{-1}}\right)^{4/9}\left({y\over 0.15}\right)^{2/9}\Omega_b^{-10/9}\Omega^{1/3}.
\end{equation}
The constraints on CMB distortions improved slowly. Illarionov \& Sunyaev \citep{IllarionovSunyaev1975b} estimated an observational upper limit in the range $y<0.04$--0.1. Field \& Perrenod \citep{FieldPerrenod1977} estimated a 90\% confidence interval $0.013<y<0.05$ primarily based on the balloon measurements by \citep{Woody1975}. This value of $y$ was shown to be consistent with the X-ray background being bremsstrahlung emission from a hot ($T\simeq 4.4\times 10^8\,$K), dense ($\Omega_b\simeq 0.46$) IGM. The statistical analysis by \citep{DaneseDeZotti1978} yielded $y< 0.05$ (excluding the data by \citep{Woody1975}) and $\mu<0.01$.

New support to the case for a relatively large comptonization distortion \citep[$y=0.02\pm 0.002$;][]{Smoot1988} was provided by the sounding rocket measurements by \citep{Matsumoto1988}, although the measurements of the balloon-borne experiment by \citep{Peterson1985} did not detect significant deviations from a black-body spectrum at wavelengths down to 0.9 mm.

This was the situation close to the COBE launch date (November 18th, 1989) and the appearance of the first FIRAS results \citep{Mather1990} that have revolutionized the field. Indications of comptonization distortions were rejected and constraints on distortion parameters were improved by three orders of magnitude \citep[$|y|<1.5\cdot 10^{-5}$ and $|\mu_0|<9\cdot 10^{-5}$;][]{Fixsen1996}.

Subsequent work, mostly by Giorgio Sironi's group, concerned long wavelengths, outside the FIRAS range. The main improvements were on free-free distortions, while those on $\mu$ were only marginal. The latest constraints at 95\% C.L. are \citep{Gervasi2008} $|\mu| < 6\times 10^{-5}$ and $ -6.3 \times 10^{-6} < Y_{\rm ff} < 12.6 \times 10^{-6}$ where $Y_{\rm ff}$ is the optical depth to free-free emission, yielding an excess antenna temperature given by:
\begin{equation}
\Delta T_{\rm ff} = T_{\rm CMB} \ \frac{Y_{\rm ff}}{x^2} \label{dTFF},
\end{equation}
i.e. $-0.15 < \Delta T_{\rm ff}/{\rm K} < 0.31$ or $\delta I_{\rm ff}(600\,{\rm MHz})< 3.4\times 10^{-3}\,\hbox{MJy}\,\hbox{sr}^{-1}$.

Weaker constraints were set by the ARCADE\,2 measurements; \citep{Seiffert2011} argue that the constraints by \citep{Gervasi2008} are likely too optimistic because the fit to the data was too constrained. With the tightest present contraints [$|y|<1.5\cdot 10^{-5}$ and $\Delta T_{\rm ff}(600{\rm MHz})< 0.31\,$K] the argument by \citep{ZeldovichSunyaev1969} gives $z_{\rm max}< 37\Omega_b^{-2/3}\Omega^{1/3}(10^4\,{\rm K}/T_e)^{2/3}$, showing that current sensitivities to both the $y$ parameter and to the free-free emission of the re-ionized IGM do not reach the levels expected from our current understanding of the re-ionization history.

\begin{figure*}
\includegraphics[width=0.48\textwidth, angle=0]{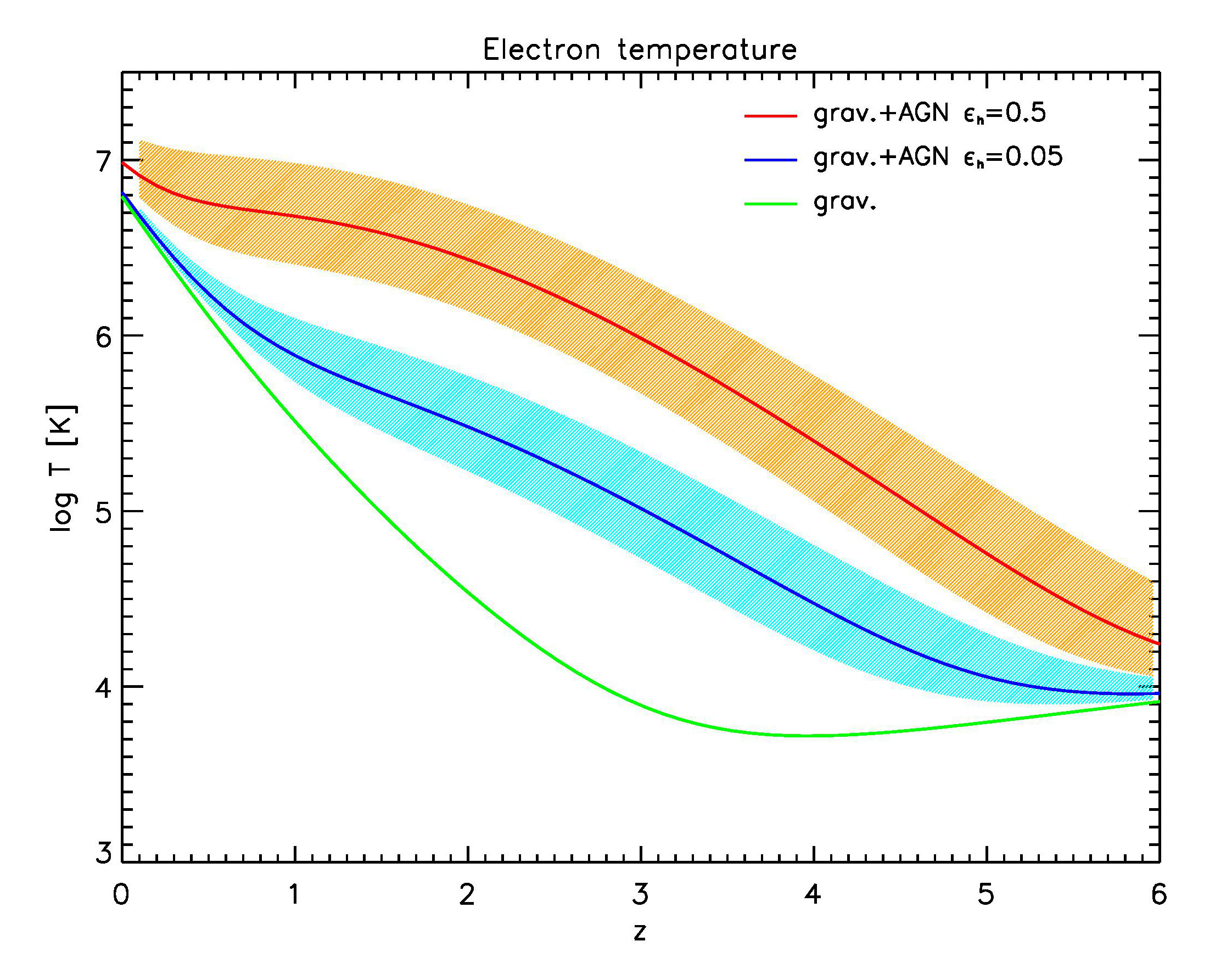}
\includegraphics[width=0.48\textwidth, angle=0]{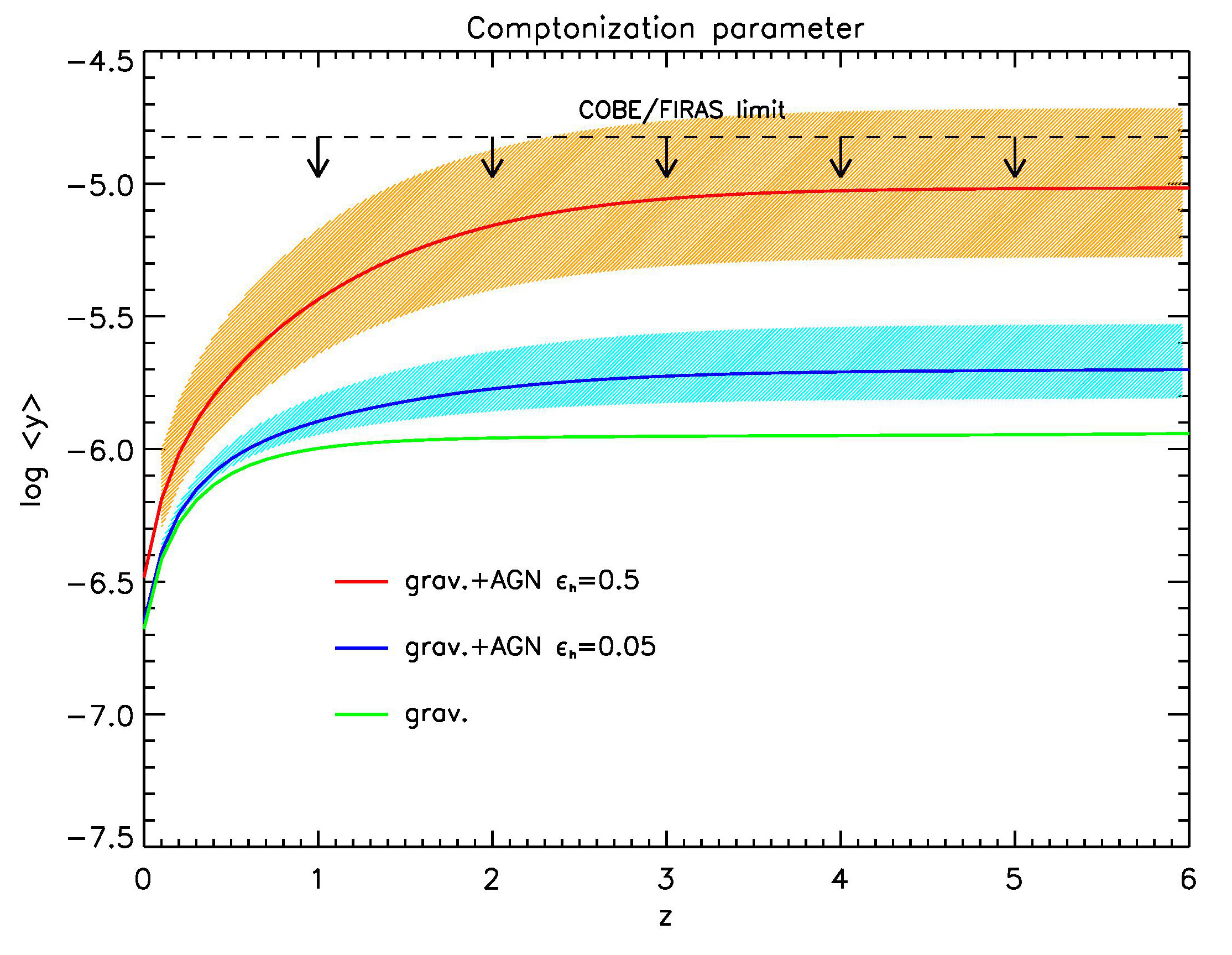}
\caption{\textbf{Left-hand panel:} evolution of the IGM temperature, $T_e$, by effect of AGN feedback heating for the standard value of the heating efficiency (solid blue line) and for the higher efficiency advocated by \citep[][solid red line]{Ruan2015}. The shaded areas around the two lines represent the uncertainties, mostly induced by those on the redshift-dependent AGN luminosity density. The green line shows the evolution of $T_e$ due to heating by shocks associated to structure formation. \textbf{Right-hand panel:} comptonization parameter $y$ yielded by the IGM heated at the temperatures shown in the left-hand panel, as a function of redshift. Essentially all the contribution to $y$ comes from $z\simlt 3$. The horizontal dashed line corresponds to the 95\% confidence upper limit from COBE/FIRAS.
 }
 \label{fig:feedback}
\end{figure*}

\subsection{Re-ionization distortions}\label{sect:reionization}

The most recent determination of the electron scattering optical depth due to re-ionization is $\tau_{\rm es}=0.066\pm 0.016$ \citep{PlanckParameters2015}. Then:
\begin{equation}
y_{\rm reion}= \int n_e \sigma_T c {k\,T_e\over m_e\,c^2}dt \simeq {k\,T_e\over m_e\,c^2}\tau_{\rm es} \simeq {T_e \over 5.93\times 10^9\,\hbox{K}}\tau_{\rm es}\simeq 2.2 \times 10^{-7}\left({\tau_{\rm es}\over 0.066}\right)\left({T_e\over 2\times 10^4{\rm K}}\right),
\end{equation}
where $T_e = 2\times 10^4\,{\rm K}$ corresponds to the peak of the hydrogen cooling curve \citep{SutherlandDopita1993}. The corresponding energy transfer to the CMB is
\begin{eqnarray}\label{eq:en_dens}
\Delta \epsilon &\simeq & 4y \epsilon_r\simeq 3.7\times 10^{-19}\left({\tau_{\rm es}\over 0.066}\right)\left({T_e\over 2\times 10^4{\rm K}}\right)\,\hbox{erg}\,\hbox{cm}^{-3}\nonumber \\
 & \simeq & 1.9\times 10^{-5}\epsilon_{\rm EBL} \simeq 1.2\times 10^{-4}\epsilon_{\rm AGN},
\end{eqnarray}
$\epsilon_{\rm EBL}\simeq 2\times 10^{-14}\,\hbox{erg}\,\hbox{cm}^{-3}$ \citep{Dole2006} being the radiation energy density produced by thermonuclear reactions in stars and $\epsilon_{\rm AGN}\simeq 3\times  10^{-15}\,\hbox{erg}\,\hbox{cm}^{-3}$ that produced by nuclear activity. The latter quantity has been estimated in two ways: from the energy density of the X-ray background adopting a bolometric correction of a factor of 20 and from the present day black hole mass density $\rho_{\rm BH}\simeq 4.5\times 10^5\,\hbox{M}_\odot\,\hbox{Mpc}^{-3}$ \citep{Shankar2009} via the equation
\begin{equation}\label{eq:en_AGN}
\epsilon_{\rm AGN} = {\eta_{\rm AGN}\over 1- \eta_{\rm AGN}}\rho_{\rm BH} c^2,
\end{equation}
adopting the standard matter to radiation conversion efficiency $\eta_{\rm AGN}=0.1$. Both methods give a very similar result.

Thus, a small fraction of the energy produced by nuclear reactions in stars and/or by nuclear activity coming out in mechanical form and going into heating of the IGM is enough to raise the electron temperature to values much higher than $T_e \simeq 1$--$2\times 10^4\,$K, the value expected by effect of the ionizing radiation.

Current galaxy evolution models advocate strong feedback effects from supernova explosions and from Active Galactic Nuclei (AGNs), generating superwinds. Such feedback effects are felt responsible for the differences in shape between the halo mass function and the galaxy luminosity or stellar mass functions \citep[e.g.][]{SilkMamon2012}. They may heat a substantial fraction \citep[$\sim 40$--50\% according to][]{CenOstriker2006} of the IGM to $10^5$--$10^7$\,K. The simulations by \citep{Tornatore2010}, including different feedback recipes, indicate that 90\% of the baryonic mass at $z> 3.5$ has $T$ in the range $10^4$--$10^5$\,K, but the fraction of warmer gas, with temperatures in the range $10^5$--$10^7$\,K, increases with decreasing $z$, reaching $\simeq 50\%$ at $z=0$.

Most AGN feedback models imply that $\simeq 5\%$ of the AGN bolometric emission is in a mechanical form \citep{Lapi2005, Lapi2006, ZubovasKing2012}. There is ample and mounting observational evidence of AGN-driven galaxy superwinds \citep{Fabian2012}. \citep{Harrison2012} find outflow powers $\sim 0.2$--5\% of the AGN luminosities, but with large uncertainties; \citep{Genzel2014} report evidence of widespread AGN-driven ionized outflows in the most massive star-forming galaxies at $z=1$--3; \citep{Brusa2015} find lower limits to the ratio of the kinetic to bolometric luminosity of the AGN nucleus in the range 0.1--5\% for a sample of $z\sim 1.5$ X-ray selected, obscured QSOs.

According to \citep{ZubovasKing2012}, the energy-driven superwinds originated in the immediate vicinity of the active nucleus shock-heat the ISM creating a two-phase medium in which molecular species (cold phase, comprising only $\sim 10\%$ of the gas) co-exist with hot gas. The large-scale outflows generated by super-winds are capable of driving the ISM out of the galaxy potential well. Their energy is then dissipated in the IGM.

Using the redshift-dependent mean AGN luminosity density, $\rho_L(z)$ \citep[inset in Fig.~1 of][]{Aversa2015}, we estimated the mean electron temperature, $T_e$, given the fraction of $\rho_L$, $\epsilon_h$, that goes into heating of the IGM, by solving the equation:
\begin{equation}\label{eq:Te}
{{\rm d}T_e(z)\over {\rm dt}} = {\epsilon_h\rho_L(z)\over 3n_e k} -{T_e \over \tau_{\rm e \gamma}} -{2T_e \over \tau_{\rm exp}},
\end{equation}
where $\tau_{\rm e \gamma}$ is the cooling time of electrons by inverse Compton scattering, given by eq.~(\ref{eq:te_gamma}). The  last term in the right-hand side accounts for the cooling by adiabatic expansion, with timescale $\tau_{\rm exp}$ given by eq.~(\ref{eq:texp}). We have neglected the cooling by free-free emission whose rate is about two orders of magnitude lower than that of Compton cooling (see sub-Sect.~\ref{sect:timescales}).

For the standard $\epsilon_h=0.05$ we find mean electron temperatures, $T_e$, in the range $10^5$--$10^6\,$K  at $z=1$--3, when the AGN activity peaks (see Fig.~\ref{fig:feedback}). Once we have $T_e(z)$ we can compute
\begin{equation}\label{eq:yfeedback}
y_{\rm reion}\!=\! \int\! n_e \sigma_T c {k\,T_e(z)\over m_e\,c^2}dt\simeq 3.36\times 10^{-6}\int_0^{z_{\rm reion}}\!\!\!\!\!\!\! {\rm d}z \left({T_e(z)\over 10^7\,\hbox{K}}\right) {(1+z)^2 \over \sqrt{\Omega_m (1+z)^3+\Omega_{\rm rad} (1+z)^4+\Omega_\Lambda}}\,.
\end{equation}
We get $y_{\rm reion}\simeq 2\times 10^{-6}$, the main contribution coming from $z\simlt 3$.

Some data indicate substantially higher values of $y$. \citep{VanWaerbeke2014} report a correlation between gravitational lensing by large scale structure and the thermal Sunyaev-Zeldovich \citep[tSZ;][]{SunyaevZeldovich1972} effect consistent with a warm plasma of temperature $\simeq 10^6\,$K and a comoving electron number density $n_e \simeq 0.25\,\hbox{m}^{-3}$ at $z\simeq 4$. If this temperature remains constant down to $z=0$ this implies $y\ge 4\times 10^{-6}$.

\citep{Ruan2015} report the detection of a strong signal at $> 5\,\sigma$ significance by stacking tSZ Compton-y maps centered on the locations of 26,686 spectroscopic quasars. They interpret the signal as due to thermal feedback energetics amounting to 5\% of the black hole mass. In this case we obtain mean electron temperatures, $T_e$, in the range $10^6$--$10^7\,$K  at $z=1$--3 and $y\simeq 10^{-5}$, only marginally below the COBE/FIRAS upper limit. Their results, however, have been challenged by other investigations. \citep{CenSafarzadeh2015} and \citep{Verdier2015} found that the tSZ detected towards quasars in their samples can be entirely accounted for by the thermal energy of their host halos and concluded that the maximum additional feedback energy must be much smaller than that claimed by \citep{Ruan2015}.

Indications that up to ($14.5\pm 3.3)/\tau_8$ percent of the quasar radiative energy goes into heating of their environment, $\tau_8$ being their typical period of activity in units of $10^8$\,yr, were reported by \citep{Crichton2015}. \citep{Spacek2016} co-added South Pole Telescope SZ survey data around a large set of massive quiescent elliptical galaxies at $z >0.5$  and found mean angularly integrated $y$ values higher than expected from simple theoretical models that do not include AGN feedback; however the implied feedback efficiency is a bit lower than that found by \citep{Ruan2015}

An additional contribution to $y$ comes from shocks associated to structure formation. Assuming that the shocks generated by the collapse and virialization process efficiently heats up the baryons \citep[e.g.][]{MossScott2009} we get $y\simeq 10^{-6}$, in agreement with earlier estimates \citep{NathSilk2001,Refregier2000}. \citep{Hill2015} obtained $y\simeq 1.8\times 10^{-6}$, with  the dominant contribution ($y\simeq 1.6\times 10^{-6}$) coming from the sky-averaged thermal tSZ effect from the hot gas in the intracluster medium of galaxy groups and clusters. This contribution, powered by gravitational energy, is not included in our estimate, that takes into account only the heating by AGN feedback, and adds to it.

We conclude that $y_{\rm reion}\simeq \hbox{several}\times 10^{-6}$, i.e. not much below the current upper limit, may be expected. \citep{Dolag2015} used state-of-the-art cosmological hydrodynamical simulations to predict the mean \textit{fluctuating} Compton $y$ value. For the \textit{Planck} 2015 values of cosmological parameters they found $\bar{y}_{\rm fluct} = 1.57\times 10^{-6}$, not far away from the upper limit, $\bar{y}_{\rm fluct} = 2.2\times 10^{-6}$, derived by \citep{KhatriSunyaev2015} using \textit{Planck} data.

The free-free emission associated to the post-reionization plasma is much harder to detect. As shown by Fig.~\ref{fig:foregrounds} for the case $\epsilon_h=0.05$ and a typical clumping factor $C\simeq 3$ \citep{KuhlenFaucher2012}, this emission is orders of magnitude below the other signals. For the higher value of $\epsilon_h$ the amplitude of the signal is lower since its intensity scales approximately as $T_e^{-1/2}$. In the limiting case $T_e=\hbox{constant}=10^4\,$K, which is probably already too low to sustain the observed almost complete ionization level, the free-free emission increases by only $\simeq 30\%$. The difference is small because in all cases most of the signal comes from the highest redshifts, where the density is the highest, the electron cooling is fast and the temperature is not far from $10^4\,$K. The SZ effect, that scales as $n_eT_e$, is more efficient at detecting a low-density, high temperature plasma than the free-free emission that scales as $n_e^2T_e^{-1/2}$.

A detailed investigation of the free-free emission for different cosmological reionization histories was carried out by \citep{TrombettiBurigana2014}. \citep{Ponente2011} estimated the average free-free signal from ionized gas in clusters and groups of galaxies. In these objects the emission is enhanced by the higher plasma density but decreased by its higher temperature. On the whole, the sky-averaged contribution of these objects is sub-dominant.

Additional contributions to $y$ could in principle occur before recombination. Although the recombination process is very sensitive to the abundance of ionizing photons, comptonization distortions with $y< 0.01$ have no detectable effect on the CMB power spectrum measured by \textit{Planck} \citep{ChlubaSunyaev2009}.

\begin{figure*}
\includegraphics[width=0.6\textwidth, angle=0]{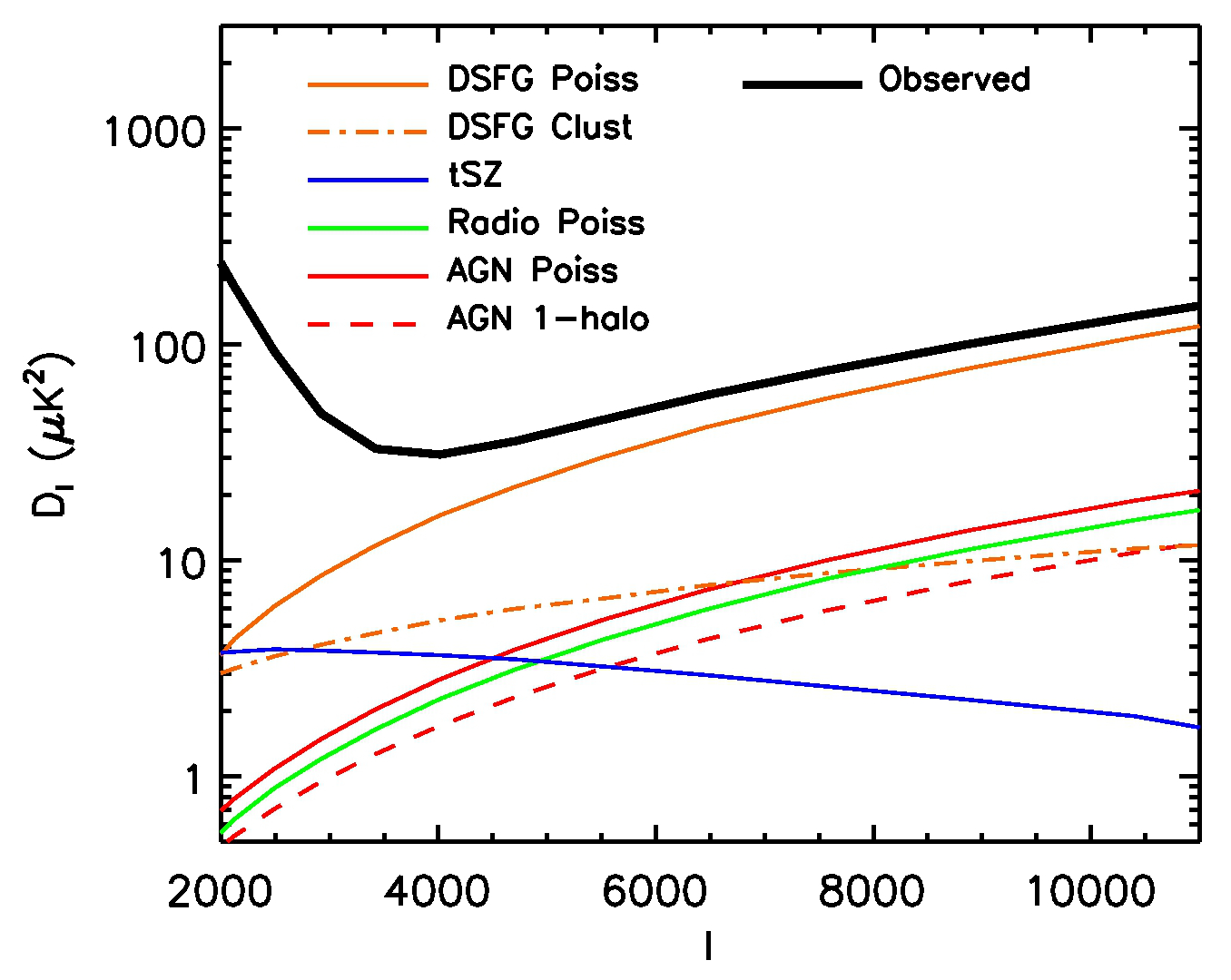}
\caption{Power spectrum of the SPT map at 150\,GHz \citep{George2015}. The thick black line is a fit to the total observed signal while the other lines are the estimated contributions from the main components: unresolved dusty star-forming galaxies (DSFG; solid and dot-dashed orange lines, for the Poisson and clustering contributions, respectively); thermal Sunyaev-Zeldovich effect due to clusters of galaxies (tSZ; solid blue line); extragalactic radio sources (Radio Poiss; solid green line). The solid and dashed red lines (AGN Poiss and AGN 1-halo) are estimates of the power spectrum of $y$--type signals due to AGN feedback in the extreme case  $\epsilon_h=0.5$ for the point source and the 1-halo models (see text). }
 \label{fig:power_spectrum}
\end{figure*}

\subsection{Power spectrum of Sunyaev-Zeldovich (SZ) fluctuations due to AGN feedback}

Since the SZ fluctuations due to AGN feedback come from discrete, although somewhat extended, sources along the line of sight, one may wonder whether the power spectrum of the signal may be a probe of it more sensitive than the spectral distortion.

The power spectrum can be computed as follows. If we assume a spherical geometry, the total thermal SZ signal of a  source at redshift $z$ is
the integral over the solid angle, $\Omega$, subtended by the source of the $y$ parameter along any line of sight, $\mathbf{\hat n}$, crossing it:
\begin{equation}\label{eq:Yz}
	Y(z) =  \int y(\mathbf{\hat n}) \, \mathrm{d} \Omega = D_{\mathrm{A}}^{-2}(z)\,\frac{\sigma_T}{m_e c^2} \int_0^{R_{\rm max}}\!\! 4\pi\, P_e(r, M, z)\, r^2 dr,
\end{equation}
where $d\Omega = dA / D_{\mathrm{A}}^2(z)$, $dA$ being an element of the source projection on the plane of the sky and $D_{\mathrm{A}}(z)$ the angular diameter distance; $P_e(r, M, z)$ is the electron pressure profile.

Because the SZ signal is the integrated pressure, the integral over the solid angle of the gas cloud provides a relatively clean measure of the total thermal energy of the gas. Setting $\hat{Y}(z)=D_{\rm A}^2(z)Y(z)$, the total thermal energy in electrons writes
\begin{equation}\label{eq:Ee}
	E_{\mathrm{e}} = \frac{3}{2} \hat{Y}(z) m_e c^2 / \sigma_T.
\end{equation}
For protons and electrons in thermal equilibrium, the total energy content of the ionized gas is simply
\begin{equation}\label{eq:Etot}
	E_{\mathrm{tot}} = (1 + \frac{1}{\mu_e}) E_{\mathrm{e}},
\end{equation}
where $\mu_e=\rho_b/(n_e m_p)$ is the mean particle weight per electron, for which we adopt the primeval value of 1.14.

According to \citep{Ruan2015} the total thermal energy due to feedback is $\simeq 5\%$ of the black hole mass energy, i.e.:
\begin{equation}\label{eq:Efeed}
	E_{\mathrm{tot}} \simeq 0.05\,M_{\rm bh}\,c^2\simeq 8.94\times 10^{60}\frac{M_{\rm bh}}{10^8\,M_\odot}\,\hbox{erg},
\end{equation}
whence
\begin{equation}\label{eq:Yzfeed}
\hat{Y}(z)=	\frac{2}{3(1 + \mu_e^{-1})} \frac{\sigma_T}{m_e c^2} E_{\mathrm{tot}} \simeq 2.61\times 10^{42}\frac{M_{\rm bh}}{10^8\,M_\odot}\,\hbox{cm}^2.
\end{equation}
As noted above, the feedback efficiency in heating the plasma advocated by \citep{Ruan2015} is a factor of 10 higher than the standard value.

The radius of the blastwave generated by the energy injected by the AGN is given by \citep{KooMcKee1992,Ruan2015}:
\begin{equation}\label{eqn:rS}
r_{\rm max}=R_{\rm BW} \approx  0.2 \, E_{62}^{1/5} t_{8}^{2/5} \left({1+z\over 2.5}\right)^{-3/5} \Delta_{100}^{-1/5} ~{\rm Mpc},
\end{equation}
where $\Delta_{100}$ is the mean enclosed pre-shock gas density in units of $100$ times the cosmological mean. Equation~(\ref{eqn:rS}) assumes a continuous energy injection over a time $t_{8}$ with luminosity $E_{62}/t_{8}$, where $E_{62}$ is the total injected energy in units of $10^{62}$\,erg, and $t_{8}$ is the time over which the feedback energy is injected in units of $10^8$\,yr. For the reference values of the parameters, the corresponding angular scale is $\theta_{\rm BW}\sim 23''$--$25''$ for $z=1$--3.

If the SZ signal is produced when the AGN is emitting at the Eddington limit, the AGN bolometric luminosity is
\begin{equation}
L_{\rm bol}=L_{\rm Edd} \simeq 1.5\times 10^{46}\frac{M_{\rm bh}}{10^8\,M_\odot}\,\hbox{erg}\,\hbox{s}^{-1}.
\end{equation}
Then in the point source approximation we have
\begin{equation}\label{eq:Poisson}
C_\ell^{P}  = \int_0^{z_\mathrm{max}}dz\frac{dV_\mathrm{c}} {dzd\Omega}\int_{L_{\mathrm{min}}}^{L_{\mathrm{max}}}dL_{\rm bol} \frac{dn(L_{\rm bol},z)}{dL_{\rm bol}} Y(L_{\rm bol},z)^2,
\end{equation}
where $dV_{\mathrm{c}}/(dz\,  d\Omega)$ is the comoving volume per unit redshift and solid angle intervals and $dn(L_{\rm bol},z)/dL_{\rm bol}$ is the comoving AGN bolometric luminosity function at the redshift $z$.

In the case of resolved sources, Poisson fluctuations decrease on angular scales inside the source area. They can be computed by summing the square of the Fourier transform of the projected SZ profile \citep[1-halo term;][]{KomatsuSeljak2002}, weighted by the number density of sources of a given BH mass and redshift:
\begin{equation}\label{eq:1h}
C_\ell^{1h}  = \int_0^{z_\mathrm{max}}dz\frac{dV_\mathrm{c}} {dzd\Omega}\int_{L_{\mathrm{min}}}^{L_{\mathrm{max}}}dL_{\rm bol} \frac{dn(L_{\rm bol},z)}{dL_{\rm bol}} \left|\tilde{y_\ell}(M_{\rm bh},z)\right|^2.
\end{equation}
The quantity $\tilde{y}_\ell(M_{\rm bh},z)$ is the 2D Fourier transform on the sphere of the 3D radial profile of the $y$-parameter of individual sources,
\begin{equation}
\tilde{y}_\ell(M_{\rm bh},z) = \frac{4 \pi r_{\mathrm{s}}^3}{D_{\mathrm{A}}(z)^2} \left( \frac{\sigma_{\mathrm{T}}}{m_{\mathrm{e}}c^{2}}\right) \int_{0}^{\infty} \ dx \ x^{2} P_{\mathrm{e}} (M_{\rm bh},z,x) \frac{\sin(\ell {x}/\ell_{\mathrm{s}})}{\ell {x}/\ell_{\mathrm{s}}},
\end{equation}
where $x=r/r_{\mathrm{s}}$, $\ell_{\mathrm{s}} = D_{\mathrm{A}}(z)/r_{\mathrm{s}}$, $r_{\mathrm{s}}$ is the scale radius of the 3D pressure profile and $D_{\mathrm{A}}(z)$ is the angular diameter distance to redshift $z$. We adopt the ``universal'' electron pressure profile proposed by \citep{Nagai2007}:
\begin{equation}
P_e(x)={P_0\over (c_{500}x)^\gamma[1+(c_{500}x)^\alpha]^{(\beta-\gamma)/\alpha}}
\end{equation}
with the best fit values of the parameters determined by \citep{Arnaud2010}: $c_{500}=1.177$, $\alpha=1.051$, $\beta=5.4905$, $\gamma=0.3081$. The coefficient $P_0$ is obtained from the condition [see eqs.~(\ref{eq:Yz}), (\ref{eq:Ee}), (\ref{eq:Etot}), (\ref{eq:Efeed}) and (\ref{eq:Yzfeed})]:
\begin{eqnarray}\label{eq:norm}
\hat{Y}(z) &=&  P_0 r_{\mathrm{s}}^3\,\frac{\sigma_T}{m_e c^2} 4\pi\, \int_0^{x_{\rm max}}
{x^2\, dx \over (c_{500}x)^\gamma[1+(c_{500}x)^\alpha]^{(\beta-\gamma)/\alpha}} \nonumber \\ &\simeq& 1.02\times 10^{-17}P_0 r_{\mathrm{s}}^3\int_0^{x_{\rm max}} {x^2\, dx \over (c_{500}x)^\gamma[1+(c_{500}x)^\alpha]^{(\beta-\gamma)/\alpha}}  \\
&=&\frac{2}{3(1 + 1/\mu_e)} \frac{\sigma_T}{m_e c^2} E_{\rm tot}\simeq 2.61\times 10^{42}\frac{\epsilon_h}{0.05}\frac{M_{\rm bh}}{10^8\,M_\odot}\,\hbox{cm}^2. \nonumber
\end{eqnarray}
Following \citep{Arnaud2010} we adopt $x_{\rm max}=5$. The corresponding radius $r_{\rm max}=5 r_s$ is set equal to the radius of the blastwave generated by the energy injected by the AGN, given by eq.~(\ref{eqn:rS}) with $\Delta_{100}=1$, $t_{8}=1$, and  $E_{62}$ is taken to be equal to $E_{\rm tot}$ in units of $10^{62}$\,erg. For the adopted pressure profile, the integral in eq.~(\ref{eq:norm}) is $=0.044$, so that
\begin{equation}\label{eq:P0}
P_0 r_{\mathrm{s}}^3 \simeq 0.65 E_{\mathrm{tot}} \simeq 0.0325\,M_{\rm bh}\,c^2 \simeq 5.81\times 10^{60}\frac{\epsilon_h}{0.05}\frac{M_{\rm bh}}{10^8\,M_\odot}\,\hbox{erg}.
\end{equation}
Once the power spectra in terms of the integrated comptonization parameter $Y(z)$ have been computed [eqs.~(\ref{eq:Poisson}) and (\ref{eq:1h})] they can be straightforwardly translated in terms of temperature or of intensity using
\begin{equation}
   \frac{\Delta T_{\rm SZE}}{T_{\rm CMB}} = Y(z) \left(x \frac{e^x+1}{e^x-1} -4\right)
   \label{eq:deltaT1}
\end{equation}
or
\begin{equation}
\Delta I_{SZE} = I_0 Y(z) \frac{x^4 e^x}{(e^x-1)^2} \left(x \frac{e^x + 1}{e^x - 1} - 4 \right)  ,
\label{eq:sze_intensity}
\end{equation}
with $x \equiv h\nu/k T_{CMB}$ and
\begin{equation}
I_0 = 2 {(k T_{CMB})^3 \over (h c)^2}\simeq 2.699\times 10^{-15}\,\hbox{erg}\,\hbox{cm}^{-2}\,\hbox{s}^{-1}\,\hbox{Hz}^{-1}\hbox{sr}^{-1} = 269.9\,\hbox{MJy}\,\hbox{sr}^{-1}.
\end{equation}
In Fig.~\ref{fig:power_spectrum} the results for the extreme case $\epsilon_h=0.5$ (solid and dashed red curves for the point-like and the 1-halo case, respectively) are compared with South Pole Telescope \citep[SPT;][]{George2015} data at 150\,GHz. The quantity plotted in this figure
\begin{equation}
D_\ell={1 \over 2\pi} \ell(\ell+1) C_\ell \simeq {1 \over 2\pi} \ell^2 C_\ell ,
\end{equation}
is approximately the power of temperature anisotropies per $\Delta\ln\ell=1$; it is frequently used in place of the power spectrum $C_\ell$. Clearly, even in this extreme case, the signal due to feedback is at least one order of magnitude below the measured power spectrum and easily swamped by other contributions.

\begin{figure*}
\includegraphics[width=0.48\textwidth, angle=0]{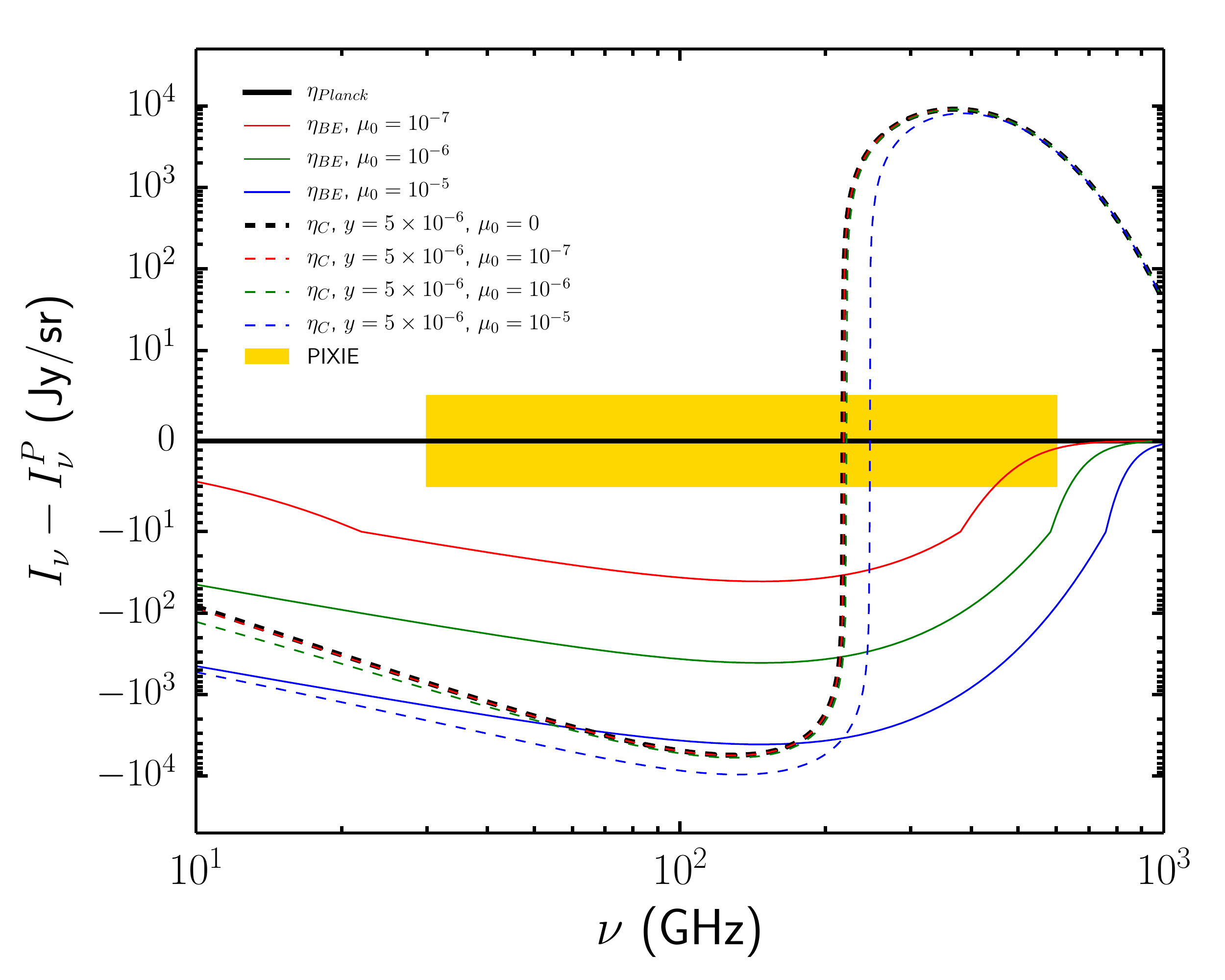}
\includegraphics[width=0.48\textwidth, angle=0]{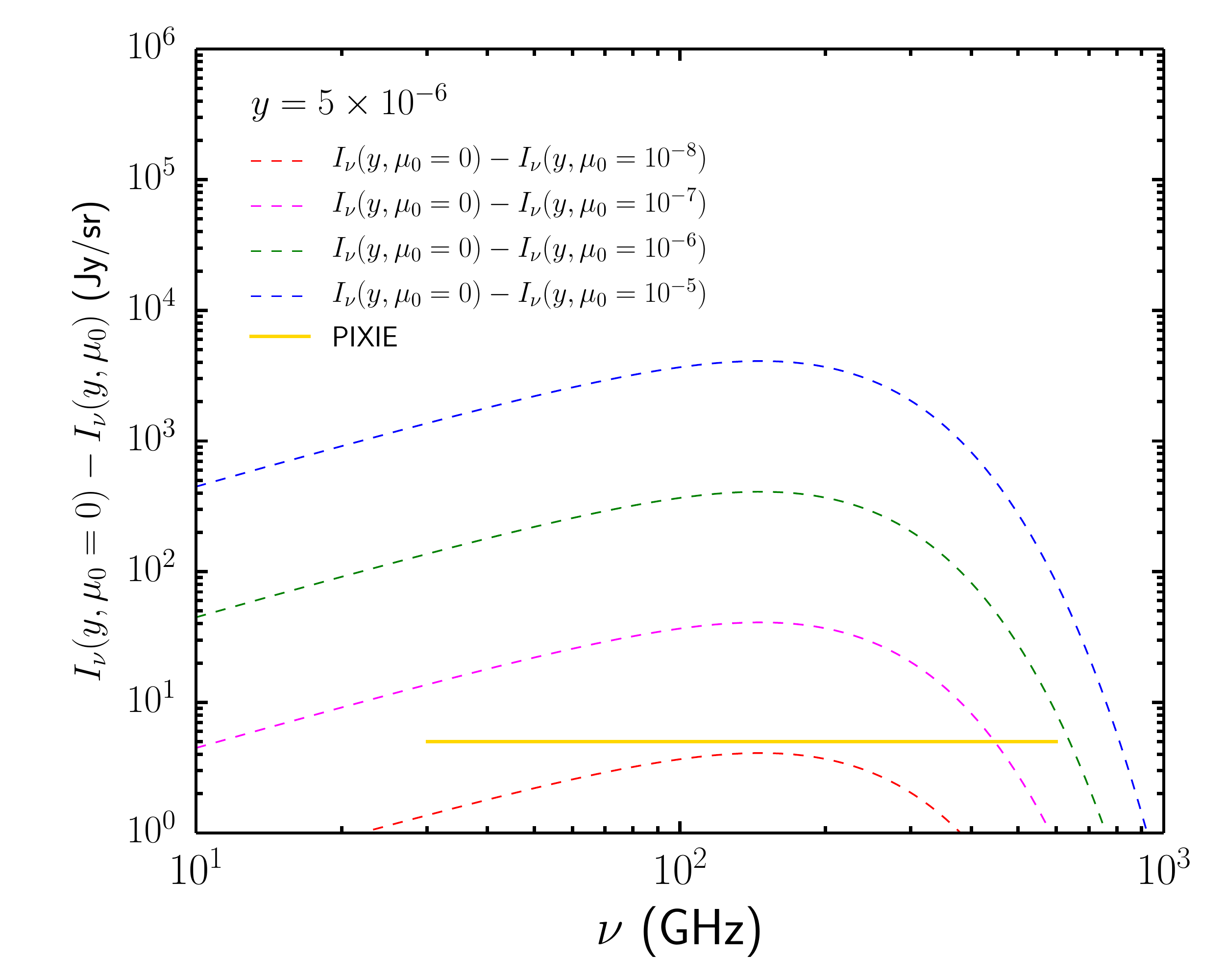}
\caption{Effect of comptonization of BE distortions during the re-ionization era. The thin dashed coloured curves in the \textbf{left-hand panel} show the comptonized BE spectra with $\mu_0=10^{-7}$, $10^{-6}$ and $10^{-5}$, and $x_c=0.01$, for $y_{\rm reion}=5\times 10^{-6}$. As $\mu_0$ decreases, the spectra obviously approach that corresponding to the comptonization with the same $y$ of an initially blackbody spectrum (thick dashed black line). The thin coloured lines show the pure BE spectra for the same values of $\mu_0$. The shaded yellow area represents the PIXIE $1\,\sigma$ sensitivity. As illustrated by the \textbf{right-hand panel}, PIXIE, whose $1\,\sigma$ sensitivity is shown by yellow line, can detect the difference from a purely comptonized spectrum down to $\mu_0=\hbox{few}\times 10^{-8}$.  }
 \label{fig:comptonBE}
\end{figure*}

\subsection{Comptonization of primordial distorted spectra}

The re-ionization distortions discussed in Sect.~\ref{sect:reionization} refer to an initially black-body spectrum. However, as mentioned above, distortions produced before recombination must have been necessarily present at the onset of re-ionization. Early weak comptonization ($y$--type) distortions are predicted to be much smaller than those due to re-ionization. Hence they will probably be swamped since, in the case of electron heating, the final distorted spectrum depends only on the total value of $y$, sum of all contributions.

In the re-ionization epoch, when electrons and radiation are no longer in thermal equilibrium, a spectrum distorted at early times acquires a $y$--type is component. In the case of a BE distortion the resulting spectrum, $\eta_{\rm BE,C}$, is described by  eq.~(\ref{eq:ydist}) with $\eta_i(x,0)$ given by eqs.~(\ref{eq:etaBE}) and (\ref{eq:mu_xe}). To first order in $u$ [eq.~(\ref{eq:u})] and for $ux^2\ll 1$ $\eta_{\rm BE,C}$ takes a form analogous to eq.~(\ref{eq:compt}):
\begin{equation}\label{eq:compt_reion}
\eta_{\rm BE,C}(x_{\rm BE},u)=\eta_{\rm BE}(x_{\rm BE})\left[1+u\,x_{\rm BE}\,\exp(x_{\rm BE}+\mu)\,\eta_{\rm BE}(x_{\rm BE})\left(x_{\rm BE}{\exp(x_{\rm BE}+\mu)+1\over \exp(x_{\rm BE}+\mu)-1} -4 \right)\right],
\end{equation}
where $x_{\rm BE}=h\nu/k T_{\rm BE}=(1-0.456\mu)x$, with $x=h\nu/k T_r$, $T_r=(1-0.456\mu)T_{\rm BE}$ being the temperature of a black-body spectrum with the same photon density as the BE spectrum. Note that, in eq.~(\ref{eq:u}), $T_r/T_e\ll 1$, so that $u\simeq y$.

The deviation from a black-body spectrum at temperature $T_r$, to first order in both $u$ and $\mu$, for $x\gg x_0$ (i.e. neglecting the frequency dependence of $\mu$) and neglecting the term of order $u\times \mu$, is then
\begin{equation}\label{eq:Cdist_reion}
\eta_{\rm BE,C}(x_{\rm BE},u)-\eta_{\rm P}(x)=\eta_{\rm P}^2(x)\,\exp(x)\left[ux \left(x{\exp(x)+1\over \exp(x)-1}-4\right)-\mu(1-0.456\mu)\right].
\end{equation}
Thus pure $\mu$--type spectra are not to be expected. The right-hand panel of Fig.~\ref{fig:comptonBE} shows that the PIXIE sensitivity would allow us to detect the difference with a purely comptonized spectrum down to $\mu_0=\hbox{few}\times 10^{-8}$.

\begin{figure*}
\includegraphics[width=0.48\textwidth, angle=0]{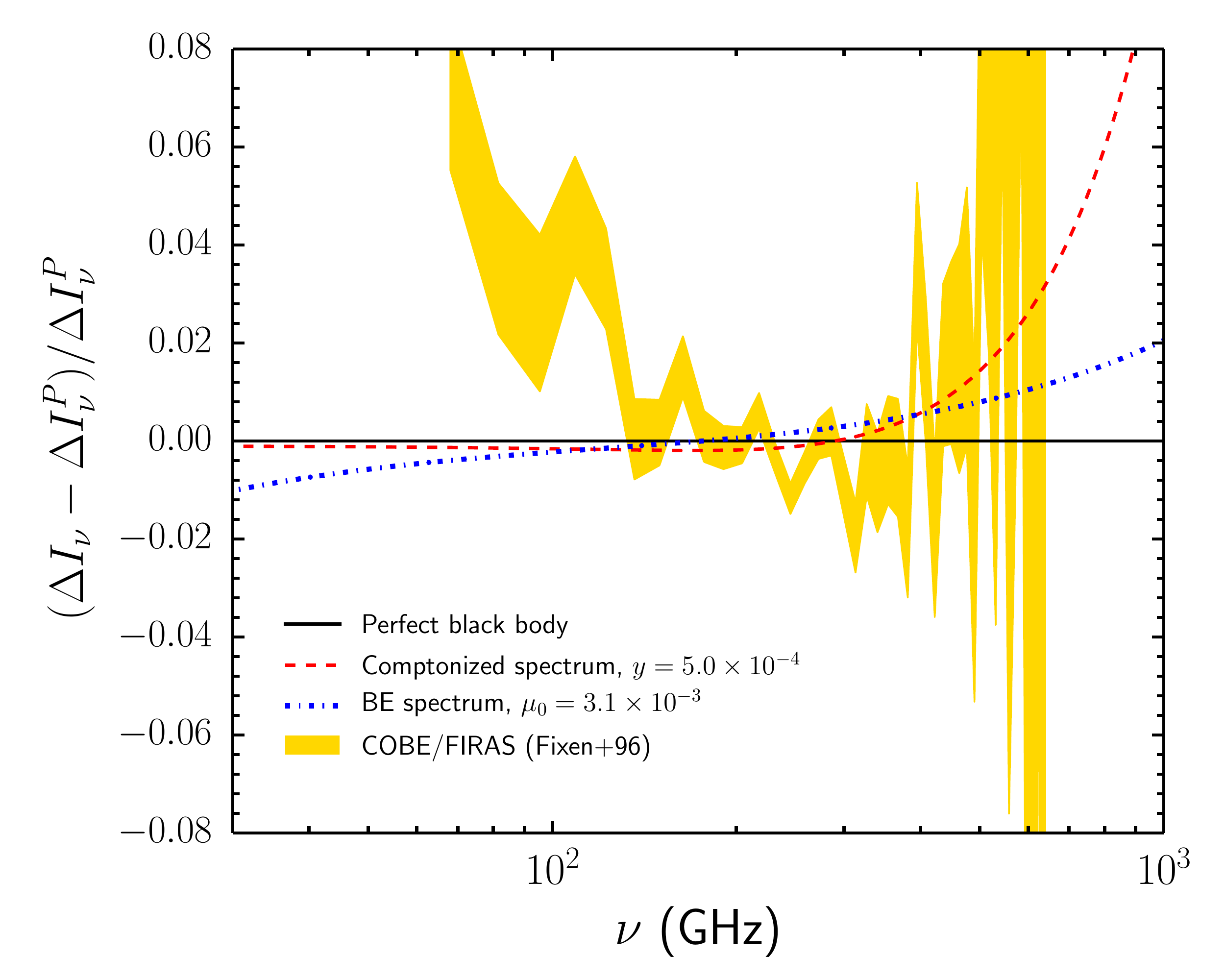}
\includegraphics[width=0.48\textwidth, angle=0]{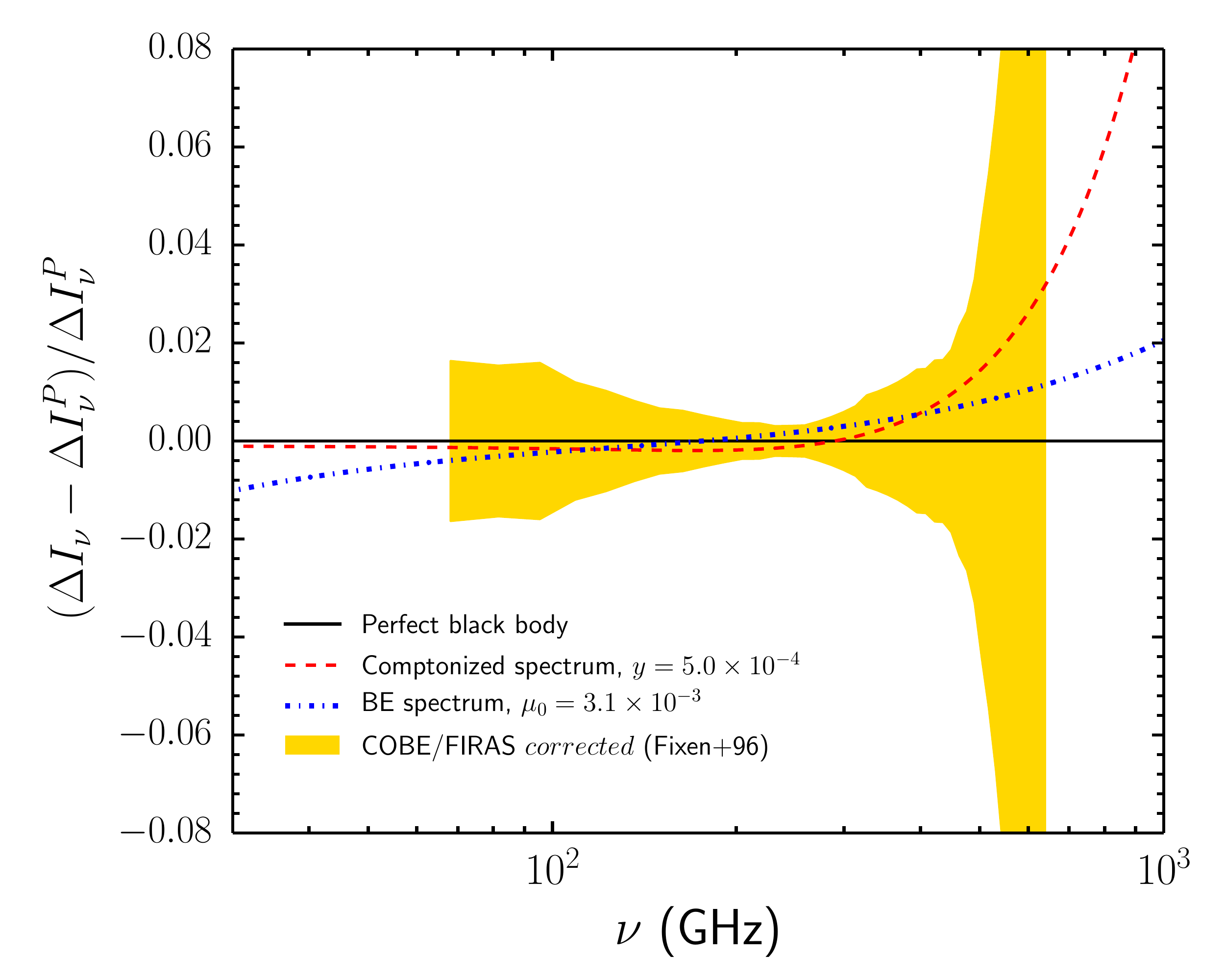}
\caption{Comparison of the dipole frequency spectra for $y$-- and $\mu$--type distortions with the perfect black-body case. In the left-hand panel, the shaded yellow region shows the dipole amplitude measurements by COBE/FIRAS \citep{Fixsen1996}. There are offsets from the black-body case likely due to systematics. If such offsets are removed (right-hand panel) the data imply 95\% confidence upper limits of $y=5.0\times 10^{-4}$ and $\mu_0=3.1\times 10^{-3}$. The dipole frequency spectra corresponding to such maximum distortions are represented, in both panels, by the dashed red line and by the dot-dashed blue line, respectively. }
 \label{fig:dipole}
\end{figure*}

\begin{figure*}
\includegraphics[width=0.48\textwidth, angle=0]{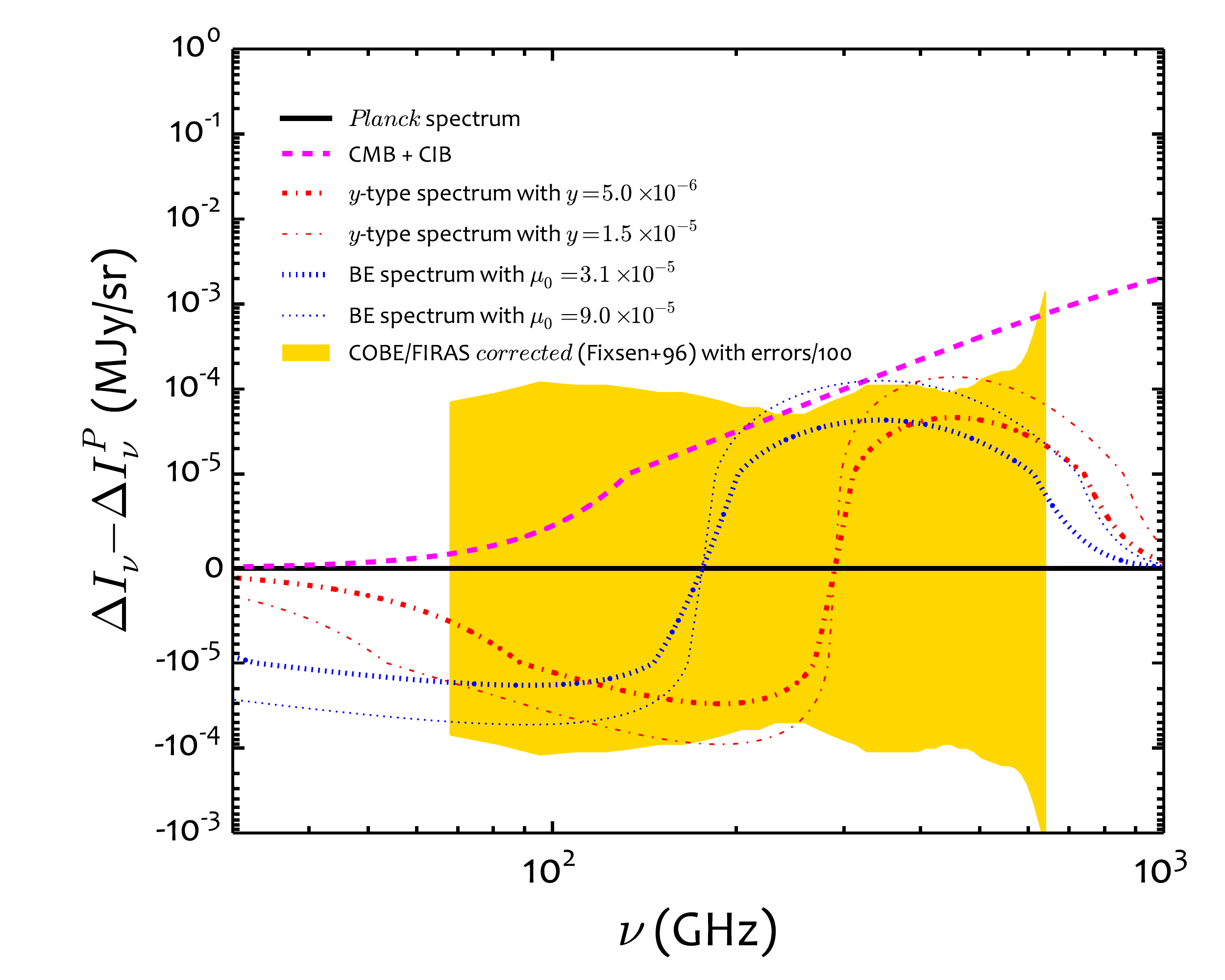}
\includegraphics[width=0.48\textwidth, angle=0]{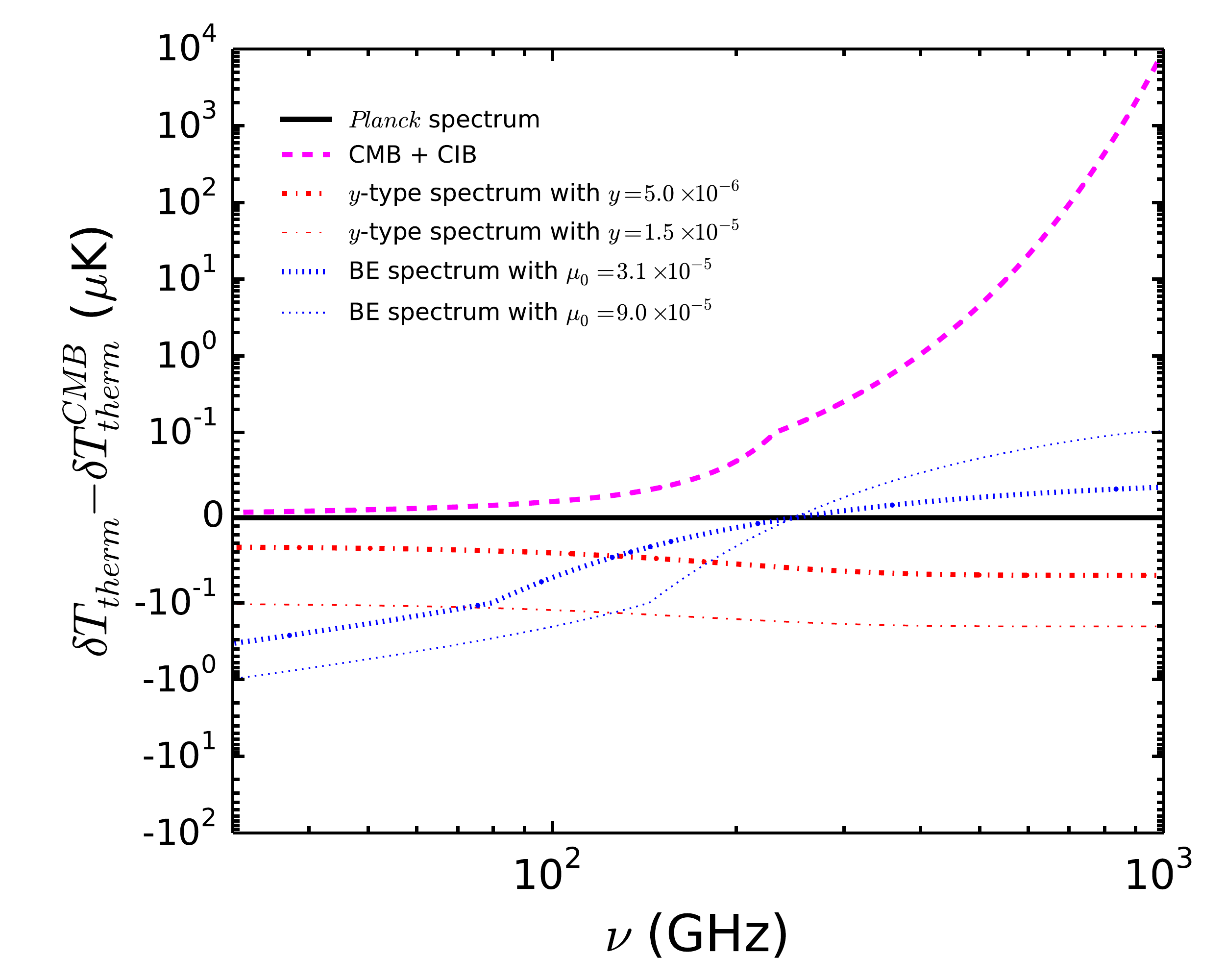}
\caption{Deviations of the dipole frequency spectrum from the black-body case in the presence of spectral distortions, in terms of intensity (\textbf{left-hand panel}) and of thermodynamic temperature (\textbf{right-hand panel}). The distorted dipole spectra correspond to the 95\% confidence upper limits ($\mu_0=9\times 10^{-5}$, $y=1.5\times 10^{-5}$) derived from COBE/FIRAS absolute spectral measurements and to those ($\mu_0=3.1\times 10^{-5}$, $y=5\times 10^{-6}$) derived assuming statistical errors on the dipole amplitude spectrum 100 times smaller than the COBE/FIRAS ones and negligible systematic errors (shaded yellow area).  The dashed magenta lines show the spectrum of the CIB dipole in terms of intensity (left-hand panel) or of thermodynamic temperature (right-hand panel) measured in maps comprising both the CMB and the CIB, after subtracting the CMB dipole. For the CIB we adopted the intensity spectrum of eq.~(\ref{eq:CIB}).
 }
 \label{fig:dipole_small}
\end{figure*}

\section{Constraints on distortions from dipole and SZ spectra}\label{sect:dipole}

\subsection{Frequency spectrum of the dipole amplitude}

The severe challenge to any attempt at improving by large factors the constraints on distortion parameters represented by foreground contamination motivates the search for methods alternative to absolute measurements and less affected by foregrounds, or at least affected by them in a different way. Two techniques have been proposed to this effect.  They are based on measurements of the spectral dependence of the dipole anisotropy and of the tSZ effect.

The dipole amplitude is directly proportional to the first derivative of the photon occupation number  or of the CMB intensity \citep{DaneseDeZotti1981,Balashev2015}. This can be easily derived from the Lorentz invariance of the photon distribution function \citep{Forman1970}: $\eta'(\nu')=\eta(\nu)$ where the primes refer to the CMB rest frame and the unprimed quantities are those measured by the observer moving at velocity $\vec{V}$ with respect to the CMB. Setting $\beta=V/c$, the Lorentz transformation of the photon frequency writes:
\begin{equation}
\nu= {\left[1-\beta^2\right]^{1/2} \over 1- \vec{\beta}\cdot\vec{n}}\nu',
\end{equation}
where $\vec{n}$ is the direction of the line-of-sight. In the case of a black-body spectrum the moving observer sees a temperature
\begin{equation}
T(\theta)= {\left[1-\beta^2\right]^{1/2} \over 1- \beta\cos\theta}T_{\rm CMB}
\end{equation}
where $\theta$ is the angle between $\vec{V}$ and $\vec{n}$. To first order in $\beta$ the observer sees a dipole $\Delta T(\theta)=T(\theta)-T_{\rm CMB}=T_{\rm CMB}\beta\cos\theta$. The measured dipole amplitude is \citep{PlanckHFIcalibration2015} $\Delta T_{\rm dipole}=\Delta T(\theta=0)=3364.46\pm 0.006\,\hbox{(stat)} \pm 0.8\,\hbox{(sys)}\,\mu\hbox{K}$.

Note that $\Delta T_{\rm dipole}$ is close, but not exactly equal to the temperature difference between the direction of motion and that perpendicular to it: the latter gets contributions from higher order multipoles. In terms of variation of the radiation intensity we have:
\begin{equation}\label{eq:dip_series}
{\Delta I_\nu(\theta)\over I_\nu}={\eta(\nu)-\eta'(\nu)\over\eta'(\nu)}= -{d\ln\eta'(\nu)\over d\ln\nu}\,{\delta\nu(\theta)\over\nu}+{\nu^2\over 2 \eta'(\nu)} \,{d^2\eta'(\nu)\over d\nu^2}\,\left({\delta\nu(\theta)\over\nu}\right)^2,
\end{equation}
with
\begin{equation}
\delta\nu(\theta)=\nu-\nu'= \left[\beta\cos\theta+{1\over 2}\beta^2\cos(2\theta)+ O(\beta^3)\right]\nu'.
\end{equation}
Thus the second order term adds a (small) quadrupolar contribution. In the case of a black-body spectrum:
\begin{equation}
I_P(\nu)={2h\nu^3\over c^2}{1\over e^x - 1}\simeq 2.6988\times 10^{-15}{x^3\over e^x - 1}\ \hbox{erg}\,\hbox{cm}^{-2}\,\hbox{s}^{-1}\,\hbox{Hz}^{-1}\,\hbox{sr}^{-1}\simeq 269.88 x^3 \eta_{\rm P}\ \hbox{MJy}\,\hbox{sr}^{-1},
\end{equation}
and the variation of intensity measured between $\theta=0$ and $\theta=\pi/2$ is  [$\delta\nu\equiv\delta\nu(\theta=0)$]
\begin{equation}\label{eq:dipole}
{\Delta I_P(\nu)\over I_P(\nu)}= {x e^x\over e^x - 1}\,{\delta\nu\over\nu} + {x^2 e^x\over e^x - 1}\,\left({\delta\nu\over\nu}\right)^2 +O\left(\left({\delta\nu\over\nu}\right)^3\right).
\end{equation}
This equation shows that, at least for small deviations from a black-body spectrum, the second order term becomes increasingly important at high frequencies.

In the CMB community it is customary to report results in terms of thermodynamic (brightness) temperature rather than of intensity. The thermodynamic temperature is related to the photon occupation number by
\begin{equation}
T_{\rm therm}={h\nu\over k\ln(1+\eta^{-1})}.
\end{equation}
The amplitude of the dipole in terms of $T_{\rm therm}$ is \citep{DaneseDeZotti1981}:
\begin{equation}\label{eq:DeltaTtherm}
\Delta T_{\rm therm}={h\nu\over k}\left\{{1\over \ln\left[1+\eta^{-1}(\nu)\right]}-{1\over \ln\left[1+\eta^{-1}(\nu(1+\beta))\right]}   \right\}
\simeq -{x \Delta T_{\rm dipole}\over (1+\eta)\ln^2(1+\eta^{-1})}{d\ln\eta\over d\ln x}.
\end{equation}
In the case of a Bose-Einstein spectrum we have
\begin{eqnarray}
{d\ln\eta_{\rm BE}\over d\ln\nu}&=&{d\ln\eta_{\rm BE}\over d\ln x_e}=-\eta_{\rm BE}{(x_e^2+\mu(x_e) x_c)e^{x_e+\mu(x_e)}\over x_e}\\
1+\eta_{\rm BE}&=&\eta_{\rm BE} e^{x_e+\mu(x_e)} \\
\ln^2\left(1+\eta_{\rm BE}^{-1}\right)&=&\ln^2\left(e^{x_e+m(x_e)}\right)=[x_e+\mu(x_e)]^2 \\
\!\!\!\!\!\!\!\!\!\!\Delta T_{\rm therm,BE}-\Delta T_{\rm dipole}&=&\left({x_e^2+\mu(x_e) x_c\over (x_e+\mu(x_e))^2}-1\right)\Delta T_{\rm dipole}\simeq \left({x_c\over x_e^2}-{2\over x_e}\right)\mu(x_e)\Delta T_{\rm dipole},
\end{eqnarray}
where $x_e=h\nu/kT_e$, $T_e$ being the Compton equilibrium electron temperature in such radiation field, which is slightly higher than the temperature, $T_r$, of a black-body with the same photon density: $T_e\simeq T_r(1+0.456\mu)$ [see eq.~(\ref{eq:etaBE})].

In the case of a weakly-comptonized ($y$--type) spectrum
\begin{equation}\label{eq:eta_C}
\eta_{\rm C}=\eta_{\rm P} + y\,x\,e^x\eta^2_{\rm P}\left(x{e^x+1\over e^x-1}-4\right),
\end{equation}
we have
\begin{equation}
{d\eta_{\rm C}\over d x}=- \eta_{\rm P}^2\left[e^x - \delta\eta\left(1+{1\over x}-2e^x\eta_{\rm P}\right)-y\,x\,e^x(1-2\eta^2_{\rm P})  \right] .
\end{equation}
For $x\Rightarrow 0$, $\Delta T_{\rm therm,C}-\Delta T_{\rm dipole}\Rightarrow -2 \,y\,\Delta T_{\rm dipole}$.

The fractional deviations from the perfect black-body case of the frequency dependence of the dipole amplitude for $y$-- and $\mu$--type distortions are illustrated in Fig.~\ref{fig:dipole} where they are compared with COBE/FIRAS measurements of the dipole amplitude spectrum \citep{Fixsen1996}. In the case of BE distortions, the comparison is made with a black-body spectrum having the same photon density as the BE spectrum.


The left-hand panel of this figure shows that such measurements exhibit significant offsets from the black-body case, likely due to systematics (calibration problems, long-term instabilities of the instrument, backlobes, ...). These offsets make difficult to derive firm constraints on distortion parameters. Removing such offsets (right-hand panel), i.e. in the case of a hypothetical experiment with the same sensitivity as COBE/FIRAS but with much lower systematic errors, we find that the data imply 95\% confidence upper limits of $y=5.0\times 10^{-4}$ and $\mu_0=3.1\times 10^{-3}$. Although these limits are much looser than those derived from absolute spectral measurements, they provide independent confirmation that spectral distortions must be much smaller than implied by some pre-COBE experiments.

The deviations from a pure black-body dipole due to spectral distortions at the COBE-FIRAS limits ($\mu_0=9\times 10^{-5}$, $y=1.5\times 10^{-5}$) at the \textit{Planck} frequencies (30, 44, 70, 100, 143, 217, 353, 545 and 857\,GHz) are $-0.98$, $-0.63$, $-0.35$, $-0.20$, $-0.10$, $-0.02$, $0.04$, $0.075$, $0.098\,\mu$K, respectively,  for a BE spectrum and $-0.10$, $-0.11$, $-0.11$, $-0.12$, $-0.14$, $-0.17$, -0.19, -0.20, $-0.20\,\mu$K   for a comptonized spectrum.

The modern technology makes possible to improve by orders of magnitude the sensitivity of CMB anisotropy measurements. The \textit{Planck} determination of the dipole amplitude quoted above has a statistical error a factor of $\sim 700$ lower than that of COBE/FIRAS, although the systematic error is only a factor of 7.5 lower.

Future experiments (see, e.g., the PIXIE \citep{Kogut2014}, LiteBIRD \citep{LiteBIRD2014}, PRISM \citep{PRISM2014} and CORE$+$ \citep{Delabrouille2015} proposals) hold the promise of improving by large factors over \textit{Planck}. If the dipole spectrum could be measured over the CORE/FIRAS frequency range with a total error 100 times lower, we would get, in the absence of detectable spectral distortions, 95\% confidence upper limits of $\mu_0\le 2\times 10^{-5}$ and $y\le 5\times 10^{-6}$, substantially better than the COBE/FIRAS limits (see Fig.~\ref{fig:dipole_small}).

\begin{figure*}
\begin{center}
\includegraphics[width=0.6\textwidth, angle=0]{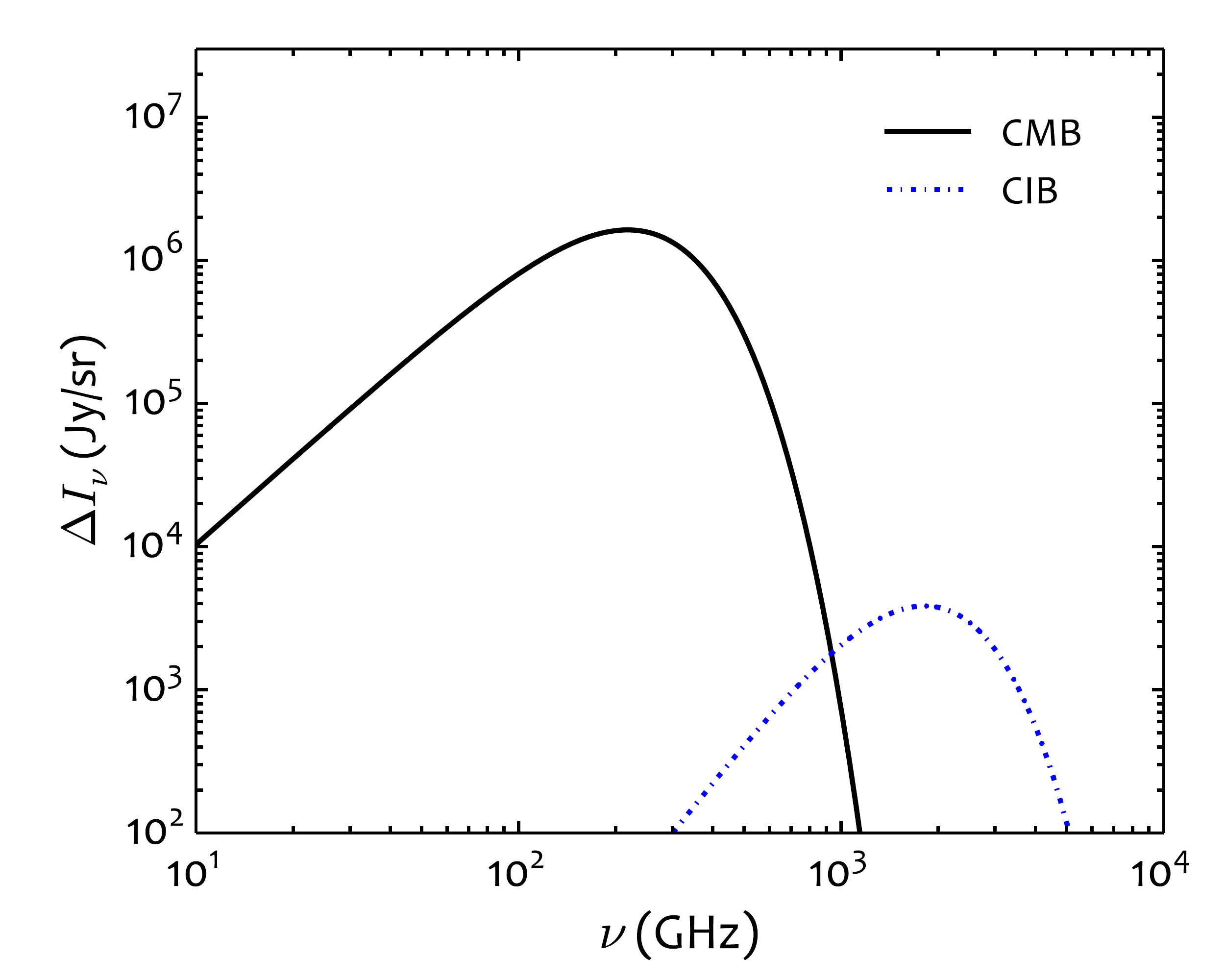}
\end{center}
\caption{Dipole intensity spectrum of the CIB compared with that of the CMB. For the CIB spectrum we have adopted the analytic description of eq.~(\ref{eq:CIB}).
 }
 \label{fig:dipoleCIB}
\end{figure*}

The figure~\ref{fig:dipole_small} also shows that the high-frequency dipole measurements may detect the CIB dipole. We may consider two possibilities, adopting in both cases the analytical description by \citep{Fixsen1998}. In the case of pure CIB maps we have
\begin{eqnarray}\label{eq:eta_CIB}
\eta&=&\eta_{\rm CIB}={c^2\over 2 h \nu^3} I_{\rm CIB}(\nu) = 1.03\times 10^{-6} {x^{k_F}_{\rm CIB}\over \exp(x_{\rm CIB})-1}\\
{d\ln\eta\over d\ln x}&=&k_F-x_{\rm CIB}{\exp(x_{\rm CIB})\over \exp(x_{\rm CIB})-1}.
\end{eqnarray}
If the map contains both the CMB and the CIB we have:
\begin{eqnarray}
\eta&=&\eta_{\rm P}+\eta_{\rm CIB}\\
{d\ln\eta\over d\ln x}&=&{x\over\eta}\left[-\eta_{\rm P}^2 e^x+{6.57\times 10^{-7}\over x^{0.36}\left(e^{x_{\rm CIB}}-1\right)}-{1.51\times 10^{-7} x^{0.64} e^{x_{\rm CIB}}\over \left(e^{x_{\rm CIB}}-1\right)^2}\right].
\end{eqnarray}
In the right-hand panel of Fig.~\ref{fig:dipole_small} we show the CIB dipole amplitude, expressed in terms of thermodynamic temperature measured in CIB$+$CMB maps, after subtraction of the CMB dipole. Clearly, at high frequencies the CIB is expected to affect the measured dipole amplitude much more than the spectral distortions compatible with the COBE/FIRAS limits. The frequency dependence of the CIB dipole intensity is shown in the left-hand panel of Fig.~\ref{fig:dipoleCIB} where that of the CMB is also shown for comparison.

If the CIB spectrum is described by eq.~(\ref{eq:eta_CIB}), we find, at the \textit{Planck} frequencies, dipole amplitudes of 0.006, 0.008, 0.013, 0.019, 0.032, 0.083, 0.53, 8.9 and $991\,\mu$K at 30, 44, 70, 100, 143, 217, 353, 545 and 857\,GHz, respectively. Thus the highest frequency \textit{Planck}/HFI maps could contain a detectable CIB signal, i.e. \textit{Planck} might provide useful constraints on the CIB intensity, currently known with a $\simeq 30\%$ uncertainty \citep{Fixsen1998}.

For completeness it is worth mentioning that \citep{Balashev2015} also pointed out, in their conclusions, the possibility of exploiting multi-frequency measurements of the dipole to gauge non-comoving foregrounds.

\subsection{Frequency spectrum of the SZ effect}

In the presence of CMB spectral distortions, the frequency spectrum of the SZ effect changes as \citep{Wright1983}
\begin{equation}
{\partial I \over \partial y} = {\partial^2 I \over \partial^2 \ln{\nu}} - 3{\partial I \over \partial \ln{\nu}}.
\end{equation}
%
The effect of $\mu$-- and $y$--type distortions compatible with COBE/FIRAS limits is small and hardly detectable. The reason for that is easily understood: the second derivative emphasizes sharp spectral features while the distorted spectra we have considered are relatively smooth. In fact the main application of this method was a test of the steep excess around the CMB peak reported by \citep{Woody1975}.

\section{Conclusions}\label{sect:conclusions}

We have reviewed some aspects of CMB spectral distortions which do not appear to have been fully explored in the literature and presented a state-of-the-art assessment of astrophysical foregrounds that may limit the accuracy of CMB spectral measurements. The strongest distortions that are necessarily present are weak comptonization, $y$--type, distortions produced by the re-ionized IGM. We have argued that, based on  recent evidences of heating of the IGM by feedback from active galactic nuclei as well as on simulations of processes associated to the formation of large scale structure, values of the $y$ parameter of $\hbox{several}\times 10^{-6}$, i.e. only a factor of a few below the COBE/FIRAS upper limit, are expected.

The re-ionized plasma also comptonizes primordial distortions adding a $y$--type contribution, so that pure $\mu$--type spectra are not to be expected. Nevertheless, the PIXIE sensitivity would allow us to detect the imprint of such distortions down to $\mu_0=\hbox{few}\times 10^{-8}$.

An assessment of Galactic and extragalactic foregrounds, taking into account the latest results from the \textit{Planck} satellite, evidences their complexity. We have shown that strong far-IR to millimeter lines (C{\sc ii}\,$157.7\,\mu$m and CO) from star forming galaxies can contribute $\sim 1\%$ of the CIB intensity over a broad frequency range. This contribution is several times higher than the PIXIE sensitivity. As a consequence, strategies for accurate CIB subtraction need to take into account the corresponding deviations from spectral smoothness. A foreground subtraction accurate enough to fully exploit the PIXIE sensitivity will be challenging, although the large number of frequency channels make PIXIE optimally suited for this purpose.

Motivated by this fact we also discussed methods to detect spectral distortions not requiring absolute measurements and showed that determinations of the frequency spectrum of the CMB dipole amplitude with an accuracy $\simeq 100$ times better than that of COBE/FIRAS could detect or constrain distortion parameters down to values about 3 times lower than the COBE/FIRAS upper limits. Such improvement in accuracy, that appears to be within the reach of planned CMB anisotropy experiments, will allow us to get close to the detection of comptonization distortions due to AGN feedback plus large-scale structure formation during the re-ionization epoch.

The CIB dipole amplitude, in thermodynamic (CMB) temperature, was estimated to be $\simeq 8.9\,\mu$K at 545\,GHz and $\simeq 991\,\mu$K at $857\,$GHz, and thus potentially detectable by accurate analyses of \textit{Planck} maps at the highest frequencies. Hence,  \textit{Planck} might provide useful constraints on the CIB dipole intensity, currently known with a $\simeq 30\%$ uncertainty.

\acknowledgments
This paper is a somewhat extended version of the invited review presented by GDZ at the conference CMB@50, at Princeton University on 10 to 12 June 2015, in celebration of 50 years of research on the Cosmic Background Radiation. GDZ warmly thanks the Organizing Committee for their kind invitation. We are most grateful to the referee, Jens Chluba, for an extremely helpful report that has led to a substantial improvement of the paper. Work supported in part by ASI/INAF agreement n. 2014-024-R.0.

\appendix
\section{Conversion formulae}\label{sect:conversion}

The background intensity $I(\nu)$ is related to the antenna temperature $T_a$ by
\begin{equation}\label{eq:Ta}
 T_a={c^2 \over 2 k \nu^2}I(\nu),
\end{equation}
so that
\begin{equation}
 {T_a\over {\rm mK}}=3.25\times 10^{-2}\left({\nu \over {\rm GHz}}\right)^{-2}{I(\nu)\over {\rm Jy/sr}},
\label{eq:Ta_mK}
\end{equation}
where $1\,\hbox{Jy}=10^{-23}\hbox{erg}\, \hbox{cm}^{-2}\, \hbox{s}^{-1}\, \hbox{Hz}^{-1}=10^{-26}\hbox{W}\, \hbox{m}^{-2}\, \hbox{Hz}^{-1}$.

The antenna temperature is related to the thermodynamic temperature $T_{\rm CMB}$ by:
\begin{equation}\label{eq:Ta_conv}
 T_a={x \over e^x -1}T_{\rm CMB},
\end{equation}
or, vice versa,
\begin{equation}\label{eq:Ttherm}
 T_{\rm CMB}={h\nu \over k\ln(x_a+1)},
\end{equation}
where $x_a=h\nu/k T_a$. Differentiating eq.~(\ref{eq:Ta_conv}) we get
\begin{equation}\label{eq:Ttherm}
\delta T_a={x^2\exp(x) \over (\exp(x)-1)^2}\delta T_{\rm CMB}
\end{equation}
or
\begin{equation}\label{eq:deltaTtherm}
\delta T_{\rm CMB} ={(\exp(x)-1)^2\over x^2\exp(x)} \delta T_a = 3.25\times 10^{-2}{(\exp(x)-1)^2\over x^2\exp(x)} \left({\nu \over {\rm GHz}}\right)^{-2}{\delta I(\nu)\over {\rm Jy/sr}}\,\hbox{mK}.
\end{equation}
For reference, PIXIE should reach a $1\sigma$ limit of $\delta I(\nu)=5\,$Jy/sr, independent of frequency in the range 30--600 GHz \citep{Kogut2014}.


\bibliography{CMB1}
\bibliographystyle{JHEP}

\end{document}